# Deformable and robust core-shell protein microcapsules templated by liquid-liquid phase separated microdroplets

**Newly added results to this manuscript: Fig. 5, Supplementary Fig. 17-20 and Movie S16-S19.**

Yufan Xu[1], Yi Shen[1], Thomas C. T. Michaels[1,3], Kevin N. Baumann[1,2], Daniele Vigolo[4], Quentin Peter[1], Yuqian Lu[1], Kadi L. Saar[1], Dominic Vella[5], Hongjia Zhu[1], Bing Li[1], He Yang[1,7], Alexander P. M. Guttenplan[1,6], Marc Rodriguez-Garcia[1], David Klenerman[1], and Tuomas P. J. Knowles[1,2]*

[1]Department of Chemistry, University of Cambridge, Cambridge, CB2 1EW, United Kingdom. [2]Cavendish Laboratory, University of Cambridge, Cambridge, CB3 0HE, United Kingdom. [3]Paulson School of Engineering and Applied Sciences, Harvard University, Cambridge, MA 02138, United States. [4]School of Chemical Engineering, University of Birmingham, Edgbaston, Birmingham, B15 2TT, United Kingdom. [5]Mathematical Institute, University of Oxford, Woodstock Rd, Oxford, OX2 6GG, United Kingdom. [6]Department of Pharmacology, University of Cambridge, Cambridge CB2 1PD, United Kingdom. [7]School of Mechanical Engineering, Hangzhou Dianzi University, No. 1158, No. 2 Street, Jianggan District, Hangzhou 310018, P.R. China. *e-mail: tpjk2@cam.ac.uk

## Abstract

Microcapsules are a key class of microscale materials with applications in areas ranging from personal care to biomedicine, and with increasing potential to act as extracellular matrix (ECM) models of hollow organs or tissues. Such capsules are conventionally generated from non-ECM materials including synthetic polymers. Here, we fabricated robust microcapsules with controllable shell thickness from physically- and enzymatically-crosslinked gelatin and achieved a core-shell architecture by exploiting a liquid-liquid phase separated aqueous dispersed phase system in a one-step microfluidic process. Microfluidic mechanical testing revealed that the mechanical robustness of thicker-shell capsules could be controlled through modulation of the shell thickness. Furthermore, the microcapsules demonstrated environmentally-responsive deformation, including buckling by osmosis and external mechanical forces. A sequential release of cargo species was obtained through the degradation of the capsules. Stability measurements showed the capsules were stable at 37 °C for more than two weeks. Finally, all-aqueous liquid-liquid phase separated and multiphase liquid-liquid phase separated systems were generated with the gel-sol transition of microgel precursors. These smart capsules are promising models of hollow biostructures, microscale drug carriers, and building blocks or compartments for active soft materials and robots.

## Key words

liquid-liquid phase separation, protein microgels, core-shell, Janus, buckling

Artificial protein-based systems can be harnessed to form hydrogels and colloids at the microscale. This approach can act as a powerful tool to investigate biomechanics of biomaterials, including shedding light on how cells and tissues behave with varying substrate mechanics (1; 2). Such studies can be crucially further advanced by extracellular matrix (ECM) protein-based materials as the components of biomimetic models *in vitro*. The mechanical properties of natural hydrogels play significant roles in modulating cellular function, but it has been challenging to track the deformation of these soft materials at the microscale (1; 3). Small and smart microgels have prospects for applications for such a purpose and more generally as environmentally-responsive carriers for catalysis, drug release, and sensing (4–8). Compared to bulk gels, spherical microgels have higher specific surface area and can thus promote more rapid exchange of substance between the microgels and environment (9; 10). Core-shell microgels present key advantages over homogeneous solid microgels, including the availability of both an outer and inner surface, and the ability to load the capsules with active ingredients. Such core-shell structures have gained increasing interest as nature-inspired constructs which can be exploited as biocompatible three-dimensional (3D) hollow scaffolds to simulate organoids or mini tissues with cavity configuration, multi-release models, hierarchical bioreactors, tailor-made cells or selective membranes to separate biomolecules statically and dynamically.

Developing core-shell microgels from proteins and their mechanical characterisation is a key enabling technology in biomaterial mechanics and regenerative medicine. Previous studies have showed the generation of core-shell non-protein microgels using two-step microfluidic techniques, 3D nested microcapillaries, or non-microfluidic methods (11–15). Progress has also been made in aqueous two-phase systems that exhibit liquid-liquid phase separation (LLPS) (**Supplementary Table 1**) as well as ECM or ECM-like systems with a high degree of biocompatibility introducing functional protein materials that are inhomogeneous (8; 9; 16–24). However, complex manufacturing methods or harsh gelation conditions of protein-based gels can limit their applications, for instance, in regenerative medicine (11; 12; 25–32). Recently, we have shown mild and versatile gelation regimes capable of producing physically- and enzymatically-crosslinked protein microgels as collagen substitutes with radial density gradients (10). Here, we report robust, deformable, and smart gelatin microcapsules produced via a one-step method in a two-dimensional (2D) microfluidic device under mild gelation conditions and their potential usage scenarios in healthcare for cargo release and controlled degradability. Specifically, the minimisation of surface energy in a liquid-liquid phase separated system can drive the self assembly of the microcapsules from ECM-substituting protein, which makes the microcapsule production scalable, accessible, and controllable for medical applications with improved biocompatibility, bioactivity, and biomimicry. All-aqueous LLPS and multiphase-LLPS systems were generated with the gel-sol transition of microgel precursors.

To fabricate the microcapsules, a 2D four-inlet microfluidic chip was used in this study (**Fig. 1a**, **Supplementary Fig. 1** and **Movie S1,S2**) (10). A gelatin/polyethylene glycol (PEG) liquid-liquid phase separating dispersed phase was chosen (phase diagram shown in **Supplementary Fig. 3**), in which the selected protein can be templated and crosslinked in a versatile and mild manner (10; 22–24). With microfluidic techniques, microcapsules with controllable shell thicknesses (E3–E5) and Janus microgels with controllable two-phase ratio (E1 and E2) were generated (**Fig. 1a–1e**). Clear interfaces between the PEG-rich and gelatin-rich phases meant these inhomogeneous microgels have sharp differences in composition (**Fig. 1b**). 3D reconstruction of confocal imaging confirmed that each layer of the microcapsules had a closed gelatin layer, while Janus microgels did not (**Fig. 1f,1g** and **Movie S3–S6**). Therefore, during demulsification, the PEG-rich phase remained encapsulated in microcapsules (sol-gel coexisting phase), but was washed off from the Janus microgels which turned into hole-shell microgels (gel phase only) (**Fig. 1b,1f** and **Supplementary Fig. 4**) (33).

Despite the use of both PEG-in-gelatin or gelatin-in-PEG flow arrangements on the chip, we found that the microcapsules (E3–E5; E8–E10) always had gelatin shells and PEG cores (**Fig. 1a** and **Supplementary Fig. 4**). This finding suggests that it is the interfacial tensions of oil/gelatin, oil/PEG, and gelatin/PEG of the microdroplets that outweigh the geometrical constraints in droplet production (**Supplementary Fig. 4**) (34). Previous studies have revealed an osmotic effect between two aqueous phase-separated phases (35); thus, varying the volume fraction $x_A$ can lead to a varying interfacial factor $(\gamma_A - \gamma_B)/\gamma_{AB}$ through water redistribution of the two aqueous phases in the microdroplets. The ultralow interfacial tension $\gamma_{AB}$ characterising the water/water interfaces would have larger impact on $(\gamma_A - \gamma_B)/\gamma_{AB}$, contributing to the Janus to core-shell transition (**Fig. 1h** and **Supplementary Fig. 5**) (21; 35–37). The configurational transition in the present study (WWO, water/water/oil) complemented previous work on the varying inhomogeneity of double emulsions (WOW, water/oil/water) (38–40). Previous studies (39) showed that eccentric core-shell structures form because of the density difference of the middle and outer phases combined with varying gelation time, which indicated that the middle/outer density mismatch could push the core off-centre with respect to the shell.

We observed that both Janus microgels and microcapsules swelled when transferred from oil to aqueous phases; their diameters increased by approximately 15–25% due to water uptake (**Fig. 1b,1d**). A larger size expansion was found in the microcapsules (E3–E5) relative to that of the hole-shell microgels (E1 and E2), a finding that highlighted the fact that the closed shells of the microcapsules could prevent the PEG leakage during demulsification (**Fig. 1b,1d**). The PEG-rich phase of hole-shell microgels dissolved during demulsification, and the swelling of hole-shell microgels depended solely on the gelatin-rich gel; in contrast, the PEG-rich cores of microcapsules stretched or expanded the gelatin-rich shells during demulsification. We found that thinner-shell microcapsules (E5) swelled more than their thicker-shell counterparts (E3 and E4); interestingly, the diameter expansion ratio was empirically linear with the flowrate ratio of gelatin to PEG solutions (**Fig. 1d** and **Supplementary Fig. 6**). One reason for this observed behaviour is that thinner-shell microgels (E5) had larger PEG cores, which meant higher capacity of water absorption. The other reason is that the thinner-shell microcapsules (E5) were less mechanically robust to the swelling of the microcapsules, compared to thicker-shell microcapsules (E3 and E4). Demulsified microgels have the potential to be used for biological applications such as drug or cell-culture studies, exploiting the structural stability of physically- and enzymatically-crosslinked protein gels (**Fig. 1b**) (10; 41).

We next extended the approach to generate such single-shell microcapsules to produce double-shell microcapsules by re-injecting the small microcapsules into a larger V-shaped flow-focusing device (**Fig. 1i** and **Supplementary Fig. 7**). These double-shell microcapsules can mimic complex natural hollow structures containing multi-compartmentalised structures, such as multi-layer ECM constructs, or could be used as multi-layer carriers for cell aggregates for artificial tissues or organs, and multi-layer reactors at small or big scales (17).

We then probed the mechanical properties of the microcapsules using a microfluidic approach. To this effect, we focused on a minority population of droplets which had multiple cores consisting of two contacting microcapsules with compressed morphologies in a single oil drop (**Fig. 2a**). We described the ellipticity of the two combining microcapsules in oil by the value $(L - W)/(L + W)$ that is related to the aspect ratio and can indicate the degree of their compression (**Supplementary information**). The encapsulation of two independent microcapsules in oil led to a reduction in the gel/oil interface area; this reduction was higher for thinner-shell microcapsules than for thicker-shell microcapsules, because more surface energy was converted into elastic strain energy of the thinner-shell capsules during their deformation (**Supplementary Fig. 10c**). We found a larger compressive deformation of the thinner-shell microcapsules (E4 and E5) relative to thicker-shell microcapsules (E3) (**Fig. 2a,2b,2c**). The ellipticity was a result of the balance between the stiffness of the two microcapsules and the water/oil and gel/oil interfacial properties; $(L - W)/(L + W)$ is expected to range from 0 to 1/3 (**Fig. 2c**). Therefore, we qualitatively conclude that thicker-shell microcapsules (E3) were stiffer than thinner-shell microcapsules (E4 and E5) (**Fig. 2c**). The encapsulation of the two microcapsules took place after the gelation of the protein shells, otherwise one bigger core-shell structure would have formed in oil (**Fig. 2a,2b**). In practice, we also found that the ellipticity value of the dual microcapsules increased after demulsification, suggesting the elastic deformation during their combination in oil or their water uptake during the demulsification into PBS.

We next probed the buckling of microcapsules by osmotic pressure and by mechanical pressure. For soft biological systems, buckling can generate dramatic elastic deformations without significant changes in extensional strain. Buckling of surfaces and capsules has attracted attention for tunable surface functionalisation, stretchable eletronics, and biomechanics; a previous study showed the thermally-induced spontaneous buckling of polydispersed core-shell microgels with non-uniform thickness which were not made with microfluidic chips (22; 42–44). In our study, we explored buckling caused by two forms of environmental stimuli, osmotic pressure gradients and direct mechanical stress.

Osmosis-induced buckling was observed after a highly-concentrated PEG solution was added to the continuous PBS phase (**Fig. 3a–3c** and **Movie S7–S9**). Confocal imaging showed that the protein-rich shells labelled by green nanospheres (GNSs) in each layer of z-stack images was smooth and closed with inner PEG-rich cores labelled by red nanospheres (RNSs), indicating that the buckling of these elastic microcapsules did not involve obvious fractures or bursts (**Fig. 3a,3b**). Buckling tended to start from the weakest or thinnest part of the microcapsules albeit their relatively uniform shell thickness (**Fig. 1b,1d**) (45). Buckling occurred rapidly within several minutes, but there was hardly any obvious recovery recorded during the 17-hour time (**Fig. 3c,3d** and **Movie S9,S10**). During the buckling, dehydration and shrinkage of the cores took place, which involved the quick transfer of water molecules from the less-concentrated PEG cores to the highly-concentrated PEG continuous phase (**Fig. 3c**). As part of this dehydration, there was a decrease in the osmotic pressure difference between the PEG-rich core and the continuous PEG phase. The buckling ended when the stiffness of the protein shells was able to counterbalance the osmotic difference (**Fig. 3c**). Previous studies showed that shrinkage, flattening, and buckling were sequentially involved in 3D latex droplets (46); and the shrinkage and a local depression at the drop surface led to both concave and convex interfaces on 2D droplets (47). The buckling of microcapsules in the present study underwent similar processes (**Fig. 3c**) (46; 47). In contrast, the recovery process mainly resulted from the water uptake of the highly-concentrated PEG-rich cores. In this process, water molecules tended to overcome the water-PEG and PEG-PEG interactions and transferred from the continuous PEG phase to the PEG cores, and thus the PEG-rich cores swelled and became more diluted (**Fig. 3d**). In practice, heating and stirring are usually needed to provide enough energy to promote the homogeneous dissolution or swelling of PEG in bulk solution. The water uptake of the dehydrated microcapsules was not an exact retraced course of the water loss of wetter microcapsules. Buckling of core-shell microcapsules was not observed in low-concentration PEG solution (**Supplementary Fig. S19).** These two processes can have different rates, as the degree of swelling of hydrogels de- pend on many factors such as network density, solvent nature, and polymer/solvent interactions (48).

Before buckling, the osmotic pressure difference outside and inside a protein shell and the tensile stress in the protein shell are balanced, and therefore the microcapsules swelled during demulsification (**Fig. 1d,3c**). The addition of highly-concentrated PEG solution to the continuous phase triggered the deformation of the microcapsules in two steps: 1) shrinking of the shell, and 2) inward buckling (**Fig. 3f**). With the ongoing shrinking step, the tensile stress in the protein shells transformed into compressive stress with increasingly compact protein shells (**Fig. 3c,3f**). When the shell became nearly incompressible, then the inward buckling step began at the weakest or thinnest part of the shells, including flattening and inward concaving (**Fig. 3c,3f**) (49; 50). The deflection, or the indentation, of the shell was perpendicular to the concentrated compressive stresses, further increased by the osmotic pressure (**Fig. 3f**). The buckling caused a concave surface and a convex surface of the microcapsule, and the work done by osmotic pressure was transformed into the strain energy stored in the microcapsule. Buckling thicker-shell microcapsules (E3) would need more energy than thinner-shell microcapsules (E4 and E5); it is well known that the energetic cost of bending a thin sheet (thickness, $t$) scales with $t^3$ (3). Bending energy and elastic energy are closely related to the conformations of the buckled invagination, and the critical external pressure required to trigger the buckling scales with $t^2/R^2$ ($R$, capsule radius) (See **Supplementary information** and **Supplementary Fig. 11** for a justification of this scaling law) (3; 49; 50). With increasing loss of the capsule volume, the bucked zone can even experience a transition from an axisymmetric dimple (primary buckling) to asymmetrical wrinkles (secondary buckling) (51; 52). Capsules with a lower bending stiffness and with a larger volume loss are more likely to develop secondary buckling (39; 52).

We next explored buckling caused by mechanical pressure in a microfluidic mechanical testing device (53). In this geometry, the oil pressure drop pushed a microcapsule into the narrow part of a V-shaped channel where deformation of the microcapsule occurred (**Fig. 3a,3e**). When a microcapsule made contact with the PDMS walls, the shell first became flattened; further compressing of the microcapsules led to increasing contact area between the shell and the PDMS walls, and the accumulated compressive stress then buckled the microcapsules (**Fig. 3g** and **Movie S11–S15**). The buckling induced by the mechanical pressure was a result of the compressive stresses (**Fig. 3e,3g** and **Supplementary Fig. 12**). After the unloading of the oil pressure, the microcapsules remained buckled, which indicated that the water drop outside the shell in the oil could not be immediately re-absorbed by the protein-rich or PEG-rich phases as previously discussed in this present study (**Fig. 3d**). Therefore, the buckling of the microcapsules was relatively permanent (**Fig. 3c,3e**). The method presented in this study therefore offers a new route for the creation of passive (mechanical force assisting) semipermeable membranes, as well as applications in lock-and-key structures for bodies and hinges or joints of biocompatible and bioactive soft robots (54; 55).

It was previously reported that, together with surface tension, buckling can keep a silk thread taut (3; 56). Recent studies show that the oil/surfactant composition could affect the morphology of touching water/oil droplets, and there is a tendency to form planar gel/gel interfaces when solid microgels aggregate in oil (17; 57). However, when two microcapsules combined in oil (**Fig. 2b** and **Supplementary Fig. 10a**), the shell/core interfaces near the shell/shell contact surfaces were slightly curved, indicating the buckling of the shells of microcapsules at the shell/shell interfaces. It is therefore technically feasible to visualise the interior deformation of the microcapsules. This buckling by interfacial tension (**Fig. 2b** and **Supplementary Fig. 10a**) is less obvious than that by osmosis or mechanical force (**Fig. 3c,3e**). Since buckling decreases the core volume, it can be inferred that water must be squeezed out of the shell and may be stored within and around the dimples created by buckling (**Fig. 2a**,**2b** and **Supplementary Fig. 10b**,**10c**).

In previous studies, the inward buckling of core-shell microgels with gradually-varied shell thickness was accompanied by triangular or other polygonal indentations (39); thin shells or layers had multiple and periodic wrinkles during buckling (3; 58). This buckling is driven by the stress induced by buckling, which can be inferred directly from changes in the shell's Gaussian curvature (**Supplementary Fig. 13**) (3). In contrast, the buckling morphologies we observe remain axisymmetric and smooth in this present study (**Fig. 3**). One reason that we do not observe polygonal instabilities might be the relatively uniform shell thickness (**Fig. 1e** and **Fig. 3**); the other possible reason is that the microcapsules are not as thin as those in other studies, and the shell elasticity plays a key role in maintaining the morphological smoothness during buckling (**Fig. 1e**) (3; 49–52). The key parameter that controls this is the Föppl-von-Kármán-number $\gamma_{FvK} = 12R^2(1-\nu^2)/t^2$, where $R$, $\nu$, and $t$ are the outer radius, Poisson ratio and thickness of the capsule, respectively (51). Previous theoretical work shared that if $\gamma_{FvK} < \gamma^c_{FvK} \sim 10^4$, then only the primary buckling is observed, not the secondary buckling event (39; 51; 52). Here, $t \approx 5$ µm, $R \approx 50$ µm (**Fig. 1d**,**1e**), so that $\gamma_{FvK} \approx 10^3 \lesssim \gamma^c_{FvK}$, the microcapsules are closer in aspect ratio $t/R$ to a tennis ball, which maintains a smooth invagination upon buckling (59).

We finally explored the potential of the capsules for multi-step release of multiple cargoes. Through the enzymatic digestion of protein shells, we observed the sequential release of GNSs from the shells and RNSs from the cores (**Fig. 4b**,**4c**). In addition, RNSs in thinner-shell capsules (E5) had a faster release than those in thicker-shell capsules (E3 and E4), because the degradation of thinner-shell capsules took place in a shorter time (**Fig. 4c** and **Supplementary Fig. 14**). **Fig. 4c** showed these release curves. ECM-based protein capsules could be digested by collagenases, trypsin, matrix metallopeptidases, or elastases in mammals (8; 60). Therefore, the principle of the sequential dual release also holds potential for personalised or nanoparticle-based medicine *in vivo*, particularly when a time-sensitive therapy requires the delivery of two drugs to target disease focuses.

Following the enzyme-induced release, we next explored temperature-induced depolymerisation of the protein shells; a previous study demonstrated rupture of the synthetic-polymer shells, for example, induced by osmotic pressure (38). Gelatin can be crosslinked enzymatically as a result of the covalent bond connecting lysine and glutamine residues with the presence of transglutaminase, or crosslinked physically because of the non-covalent interactions such as Van der Waals' forces, hydrogen bonding, hydrophobic interaction, and electrostatic interaction (10). With transglutaminase, the gelatin shells underwent first physical crosslinking and then enzymatic crosslinking, which contributed to the robust networks of peptide chains of gelatin and thus improved the thermostability of the microcapsules at 37 °C (**Fig. 4d**,**4e**). The microcapsules fabricated through physical and enzymatic crosslinking were thermostable at 37 °C for more than two weeks, and thermostable at room temperature (RT) for more than one month (**Fig. 4d**). One reason of this difference is that the crosslinked triple helices were more likely to untwist at 37 °C, leading to a larger population of random coils and the gradual weakening of the hydrogels. The other explanation is that degradation was accelerated by the action of microorganism at 37 °C (**Fig. 4d**). In contrast, we also made physically-crosslinked microcapsules without transglutaminase (**Fig. 4e** and **Supplementary Fig. 16**). After demulsification, these physically-crosslinked microcapsules remained thermostable in PBS at RT, but dissolved quickly (less than 10 min) in PBS at 37 °C (**Fig. 4e**). The gelation through only physical crosslinking was weak and reversible. A possible application taking advantage of both these crosslinking approaches could be a thermosensitive and implantable medicine (**Fig. 4e**). This can be used for personalised medicine, nanoparticle-based therapies, and artificial tissue scaffolds as asynchronous drug carriers for short-term and long-term releases *in vivo* or *in vitro*.

Our recent study demonstrated the formation of all-aqueous emulsion with microgel precursors in an aqueous macromolecular crowder (62). In this present study, we fabricated LLPS'd systems by the gel-sol transition of hole-shell and core-shell microgel precursors (physically crosslinked) in PEG crowder (**Fig. 5**). Hole-shell microgels turned spherical droplets in PEG solution when heated to 37 °C, and the coalescence of droplets in close proximity was also observed (**Fig. 5a**). Core-shell microgels buckled in PEG solution at RT, but transformed into spherical core-shell droplets at 37 °C (**Fig. 3c,5b and Supplementary Fig. 17**). The coalescence of such buckled core-shell droplets was also observed; the shells coalesced because of wetting, before the fusion of cores (**Fig. 5b**). Enzymatically crosslinked microgels, however, remained intact morphologically when heated to 37 °C, which further highlighted that enzymatically crosslinked gelatin gel would not undergo thermo-induced gel-sol transition (**Supplementary Fig. 18**). The formation of LLPS (**Fig. 5a**) or multiphase-LLPS (**Fig. 5b**) systems from microgel precursors provided a simple and facile approach to generating all-aqueous emulsion without direct on-chip generation that usually required additional setups (21,35,62-67). Previously, the fusion of core-shell structures was reported in water-oil-water double emulsion or RNA-protein hollow condensates (68,69); this present study extends the fusion phenomenon to crowder-protein-crowder all-aqueous interfaces (**Fig. 5b**). On the one hand, the formation of Janus and core-shell microgels (**Fig. 1**) was achieved by gelatin/PEG LLPS and by the gelation (sol-gel transition) of gelatin. On the other hand, such Janus and core-shell microgels can be intriguingly facilitated to fabricate all-aqueous LLPS or multiphase LLPS systems by the reversible gel-sol transition of gelatin (**Fig. 5**). Our recent study demonstrated that LLPS from solid gelatin microgel precursors in PEG solution of relatively high PEG concentration (62); similarly, in this present study, multiphase LLPS from core-shell microgel precursors took place in PEG solution of relatively high concentration (**Supplementary Fig. 19**). Higher concentration of macromolecular crowding has higher depletion force and excluded volume effect that supports LLPS or multiphase LLPS (24,62,63,70,71).

In summary, we have demonstrated a route towards the fabrication of monodisperse  core-shell gelatin microcapsules with controllable and varying shell thickness.   The spontaneous formation of the microcapsules exploits aqueous liquid-liquid phase separation. Physical and enzymatic crosslinking approaches further stabilised the protein shells. The degradable microcapsules are promising as implants or nanoparticle-based therapies for sequential release or temperature-sensitive implants. All-aqueous LLPS and multiphase LLPS systems were generated with the gel-sol transition of microgel precursors. Interfacial tension, osmosis, and mechanical pressure were utilised to understand the deformation of these microcapsules, paving the path for their applications under physical, chemical, and medical circumstances that are influenced by material mechanics.

## Methods
**Materials preparation.** Polyethylene glycol (PEG) solution (30 mg/mL) was made by dissolving PEG powder (Molecular weight 300,000; Sigma-Aldrich Co Ltd, MO, US; product of USA) in phosphate buffered saline (PBS; Oxoid Ltd, Hampshire, UK) at 50 °C with magnetic stirring for 5 h. Gelatin solution (75 mg/mL) was made by dissolving gelatin powder (Sigma-Aldrich Co Ltd, MO, US; product of Germany) in PBS at 50 °C with magnetic stirring for 2 h. Enzyme solution, i.e. transglutaminase solution (50 mg/mL), was made by dissolving transglutaminase powder (Special Ingredients Ltd, Chesterfield, UK; product of Spain) in PBS at RT for 2 h, and the solution was then filtrated with a 0.22 µm filter. Gelatin solution and enzyme solution were then kept at 4 °C and used within one week. For nanoparticle encapsulation, green nanospheres (GNSs; 200 nm, 1% solids, Fluoro-Max, Thermo Scientific, CA, US) were pre-mixed in the gelatin solution (1/100, v/v), and red nanospheres (RNSs; 100 nm, 1% solids, Fluoro-Max, Thermo Scientific, CA, US) were pre-mixed in PEG solution (1/100, v/v). Fluorosurfactant (2%, w/w) (RAN biotechnologies, MA, US) was dissolved in Fluorinert (FC-40; TM, Reg; Fluorochem, Hadfield, UK) as the continuous oil phase.

**Microgel formation.** Microfluidic devices were fabricated by soft-lithography techniques as previously reported (61). A temperature-controlled microfluidic setup was used as previously reported (10). Flow-focusing V-shaped microfluidic chips were used to make physically and enzymatically crosslinked gelatin microgels. 1) To form physically-crosslinked and enzymatically-crosslinked gelatin microgels, PEG solution, gelatin solution, enzyme solution, and continuous oil phase were loaded in four separate syringes with polythene tubings. The tubing containing gelatin phase was fixed on a hot plate at 37 °C. The flowrates of these liquids were controlled by a neMESYS pump system (CETONI GmbH, Korbussen, Germany). The inlet positions of the PEG solution and enzyme solution were also switched for different experiment geometries. The microdroplets were formed at the flow-focusing junctions of the microfluidic chips at 37 °C, and were then collected and incubated in Eppendorf tubes at room temperature (RT) overnight. The microgels were then demulsified with 10% 1H,1H,2H,2H-perfluoro-1-octanol (Sigma-Aldrich Co Ltd, MO, US) and finally rinsed in PBS (9). 2) To form physically-crosslinked gelatin microgels, PEG solution, gelatin solution, PBS, and continuous oil phase were loaded in four separate syringes with polythene tubings. Similarly, the microdroplets were formed at the flow-focusing junctions of the microfluidic chips at 37 °C, and were then collected and incubated in Eppendorf tubes at RT overnight. The microgels were then demulsified with 10% 1H,1H,2H,2H-perfluoro-1-octanol at RT or lower temperature. 3) The calculation of Janus to core-shell transition could be found in **Supplementary information**.

**Optical microscopy.** 1) The bright-field images of the formation of microdroplets were taken with a high-speed camera (MotionBLITZ EoSens Mini1-1 MC1370, Mikrotron, Unterschleissheim, Germany) on a microscope (Oberver.A1, Axio, Zeiss, Oberkochen, Germany). The dark-field fluorescent images of the microgels were taken with a CCD camera (CoolSNAP MYO, Photometrics, AZ, US) on a microscope (Oberver.A1, Axio, Zeiss, Oberkochen, Germany); for the GNSs and RNSs, a 49001 filter (excitation wavelength 426–446 nm, emission wavelength 460–500 nm) and a 49004 filter (excitation wavelength 532–557 nm, and emission wavelength 570–640 nm) were respectively used with a compact light source (HXP 120 V, Leistungselektronik Jena GmbH, Jena, Germany). 2) Confocal images were taken on a microscope (Leica TCS SP5, Germany) for GNSs (excitation wavelength 468 nm, emission wavelength 508 nm) and RNSs (excitation wavelength 542 nm, and emission wavelength 612 nm). Data were analyzed with Python and ImageJ.

**Phase diagram.** Gelatin solution (0–100mg/mL stepwise) and PEG solution (0–60mg/mL stepwise) were mixed (v/v 1:1) at 37 °C on a vortex mixer (Fisherbrand) at 2000 rpm for 20 s. Brightfield images of miscibility of the two solutions were immediately taken. Obvious water-water emulsion was considered two-phase.

**Microcapsules encapsulated in a single oil drop.** Microcapsules in oil were transferred from the Eppendorf tubes to glass slides. A minority population of microcapsules had spontaneously combined, resulting in the forms of two or more microcapsules in a single oil drop. Brightfield microscopy images were taken.

**Buckling studies.** 1) Buckling through osmotic pressure. A highly-concentrated PEG solution (60 mg/mL) was added to the demulsified microcapsule solution (v/v, 3/1) to cause the dehydration of microcapsules. Then, this PEG solution was removed, and PBS was added to the microcapsules to study the rehydration of microcapsules. 2) Buckling through mechanical forces. Microcapsules in oil were squashed into a microfluidic device (manuscript in preparation) (53); additional oil was injected into the bypass channel of the device at different flowrates. Images were were taken with a CCD camera (CoolSNAP MYO, Photometrics, AZ, US) or a high-speed camera (MotionBLITZ EoSens Mini1-1 MC1370, Mikrotron, Unterschleissheim, Germany) on a microscope (Oberver.A1, Axio, Zeiss, Oberkochen, Germany).

**Microcapsule dissolution.** 1) Trypsin (concentration 0.25%) (Life Technologies Ltd, Paisley, UK) was added to the physically- and enzymatically-crosslinked microcapsules in PBS (1/1 v/v) in a 96-well UV-transparent half area plate (Corning Incorporated, ME, US); and the dark-field fluorescent time-lapse images for GNSs and RNSs were taken with a CCD camera (CoolSNAP MYO, Photometrics, AZ, US) on a microscope (Oberver.A1, Axio, Zeiss, Oberkochen, Germany) with a 49001 filter (excitation wavelength 426–446 nm, emission wavelength 460–500 nm) and a 49004 filter (excitation wavelength 532–557 nm, emission wavelength 570–640 nm). 2D confocal imaging was also performed on a confocal microscope (Leica TCS SP5, Germany). 2) Microcapsules in PBS were incubated at 37 °C, and then their existence was checked on a microscope to determine their thermal robustness or stability with time.

**Formation of all-aqueous LLPS systems and multiphase-LLPS systems from microgel precursors.** 1) All-aqueous LLPS system. The Janus (hole-shell) microgel suspension was mixed with 6% PEG solution (v/v, 1/10) at RT. Such mixture was heated to 37 °C to form an LLPS system. 2) All-aqueous multiphase-LLPS system. The core-shell microgel suspension was mixed with 6% PEG solution (v/v, 1/10) at RT. Such mixture was heated to 37 °C to form a multiphase-LLPS system. Imaging was conducted at 37 °C in a thermostatic chamber.

## Supplementary information
**Supplementary information** is available online, with detailed methods. Codes of analyses, plots and schematics can be provided by the authors upon reasonable requests.

## Acknowledgements
The research leading to these results has received funding from the Cambridge Trust (Y.X.; B.L.), the Jardine Foundation (Y.X.), Trinity College Cambridge (Y.X.), Peterhouse College Cambridge (T.C.T.M.), the Swiss National Science foundation (T.C.T.M.), the Engineering and Physical Sciences Research Council (K.L.S.), the Schmidt Science Fellowship program in partnership with the Rhodes Trust (K.L.S.), St John's College Cambridge (K.L.S.), China Scholarship Council (H.Z.; B.L.), EPSRC Cambridge NanoDTC (EP/037221/1; A.P.M.G.), the Newman Foundation (T.P.J.K.), the Wellcome Trust (T.P.J.K.), and the European Research Council under the European Union's Seventh Framework Programme (FP7/2007-2013) through the ERC grant PhysProt (agreement n° 337969; T.P.J.K.).

## Conflict of interest
The authors declare no competing financial interest.

## Author contributions
Y.X., and T.P.J.K. conceived and designed the experiments. Y.X. developed the microgels, performed the experiments, and analysed the data. Y.S. advised on the experiments. T.C.T.M., and Y.X. developed the surface energy calculation. Y.X., and K.N.B. performed the confocal microscopy experiments. Y.X., and Q.P. studied the buckling process. D. Vella gave useful advice on the buckling studies. A.P.M.G. and K.L.S. originally designed the mechanical testing microfluidic device which was modified by Y.S., Y.L., and Y.X. in **Fig. 3e**. B.L. and D.K. assisted at the imaging for **Fig. 5**. H.Y. performed the finite element simulation. Y.X. wrote the paper, and all authors commented on the paper.

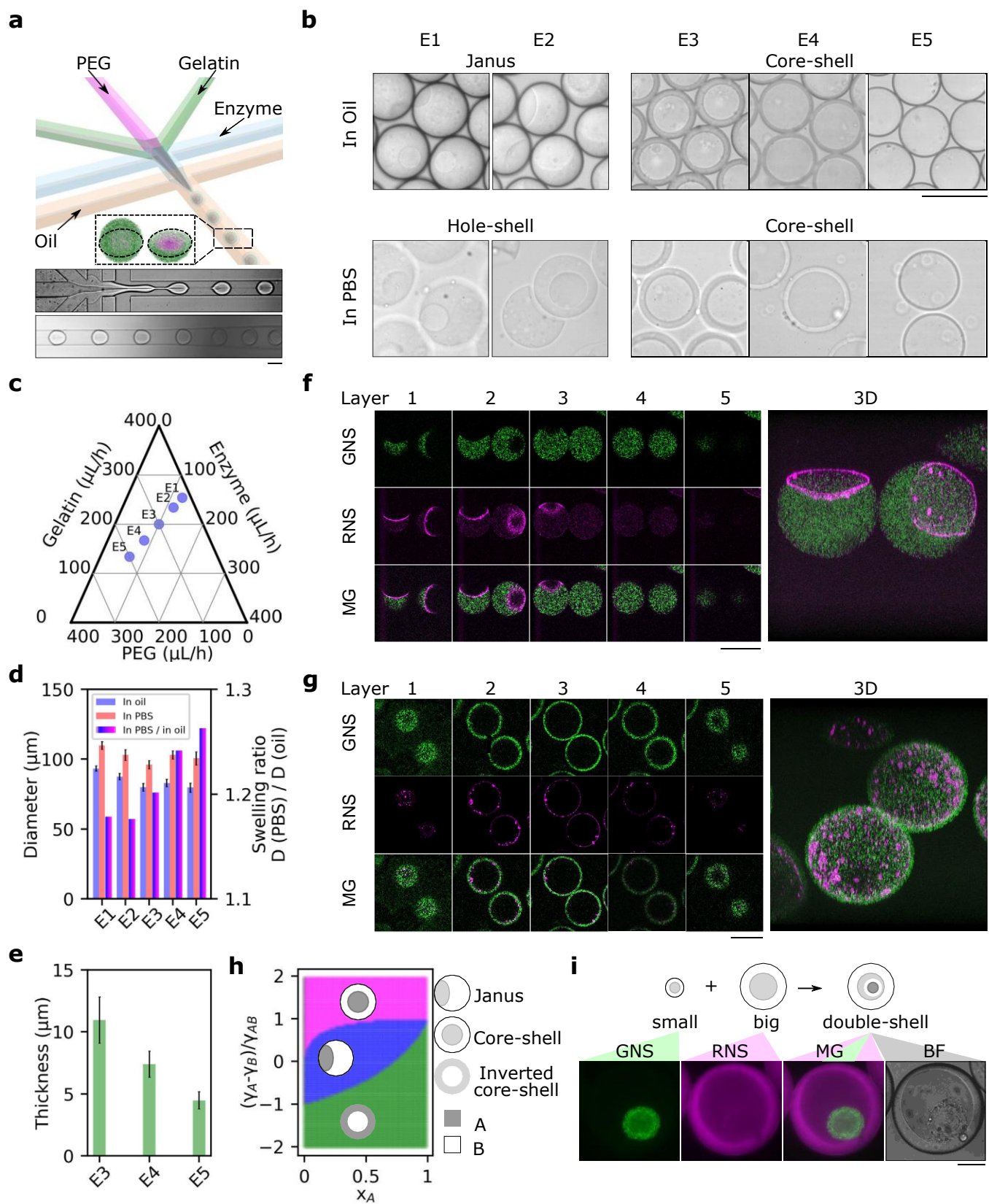

**Fig. 1 The formation of Janus microgels and microcapsules. a**, The formation of microgels on a 2D microfluidic chip. A core-shell microgel and its cross section are shown in the schematic. Scale bar, 100 μm. **b**, Five different proportions of materials lead to Janus microgels or microcapsules (core-shell microgels). Janus microgels became hole-shell microgels after demulsification. No nanospheres were added in the microgels in **b**. Scale bar, 100 μm. **c**, The flowrates of gelatin, PEG, and enzyme phases used to generate microgels shown in **b**. Continued on next page.

**d**, Left, diameter of microgels in oil and demulsified in PBS, shown with the standard deviation in the bar chart. Each sample size of E1–E5 in oil and in PBS contains 100 realisations. Right, the swelling ratio of microgels, namely the ratio of diameter after to before demulsification. **e**, Thickness of the shells of microcapsules in **b** and **d**, shown with the standard deviation. Each sample size of E3–E5 in PBS contains 100 realisations. **f** and **g**, Confocal images of hole-shell microgels (**f**) and microcapsules (**g**), layer 1-5 refers to five evenly spaced 2D layers of a 3D confocal imaging (**Movie S3**–**S6**). Green nanospheres (GNSs) and red nanospheres (RNSs) were pre-mixed in the gelatin and PEG solutions, respectively. Scale bar, 100 μm. **h**, Phase diagram illustrating the minimum energy configurations of Janus, core-shell, and inverted core-shell microgels. Here, A and B are two immiscible aqueous solutions. $x_A$ is the volume ratio of A solution in a droplet to that of the droplet. $\gamma_A$, $\gamma_B$, and $\gamma_{AB}$ are interfacial tension coefficients at the interfaces of A solution/oil, B solution/oil, and A/B solution. **i**, A double-shell microcapsule demonstrating a smaller microcapsule (GNSs in shell) encapsulated in a bigger microcapsule (RNSs in shell). Scale bar, 100 μm.

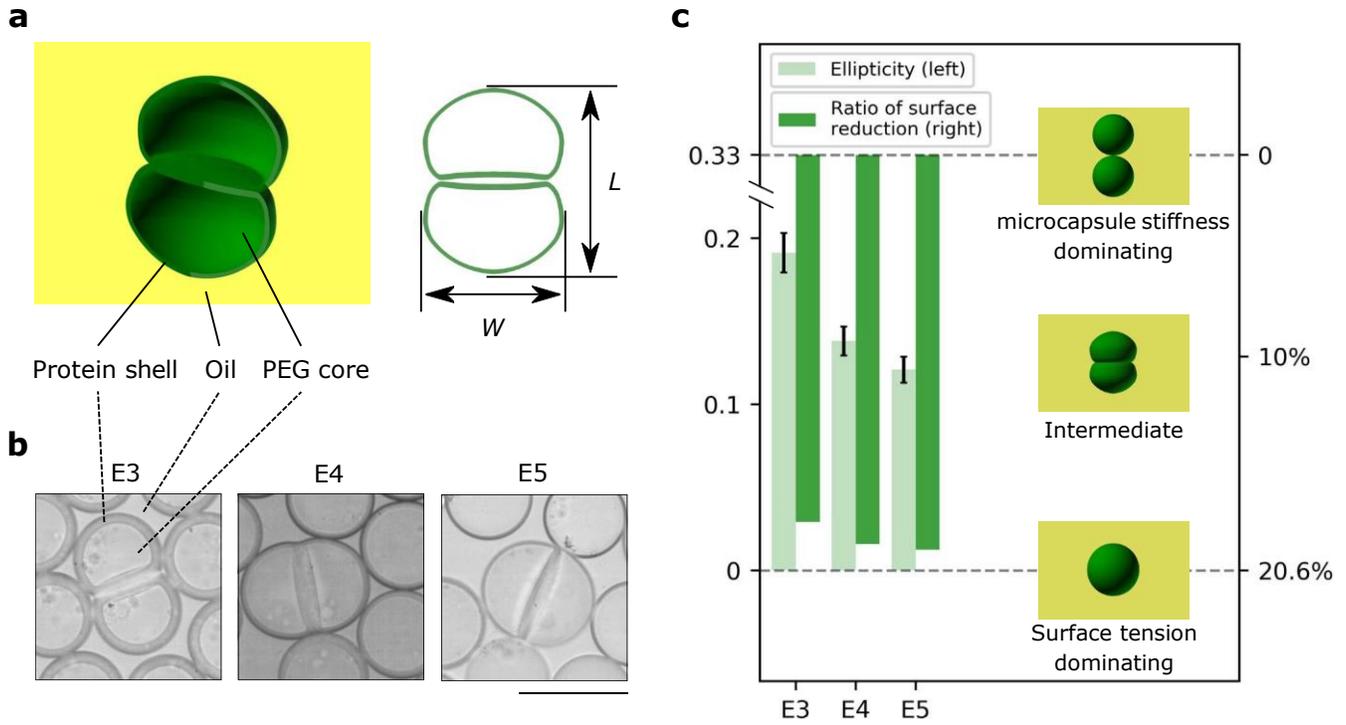

**Fig. 2 Analysis of the shapes of stressed microcapsules in oil. a,b,** Schematic (**a**) and microscopy images (**b**) of two compressed microcapsules with varying thickness in oil. *L* and *W* respectively mean the length and the width of the structures (**Supplementary Fig. 10**). Scale bar, 100 μm. **c,** The ellipticity (left y axis) defined as $(L - W)/(L + W)$ of the structures in **a,b**, shown with the standard error of the mean in the bar chart. Sample size of E3–E5 is respectively 33, 75, and 65. The ratio of reduction of the gel/oil surfaces (right y axis) of a big fused microcapsule from two small spherical microcapsules in **a,b**. The y-axis position of the intermediate schematic was at the approximate position in the bar chart.

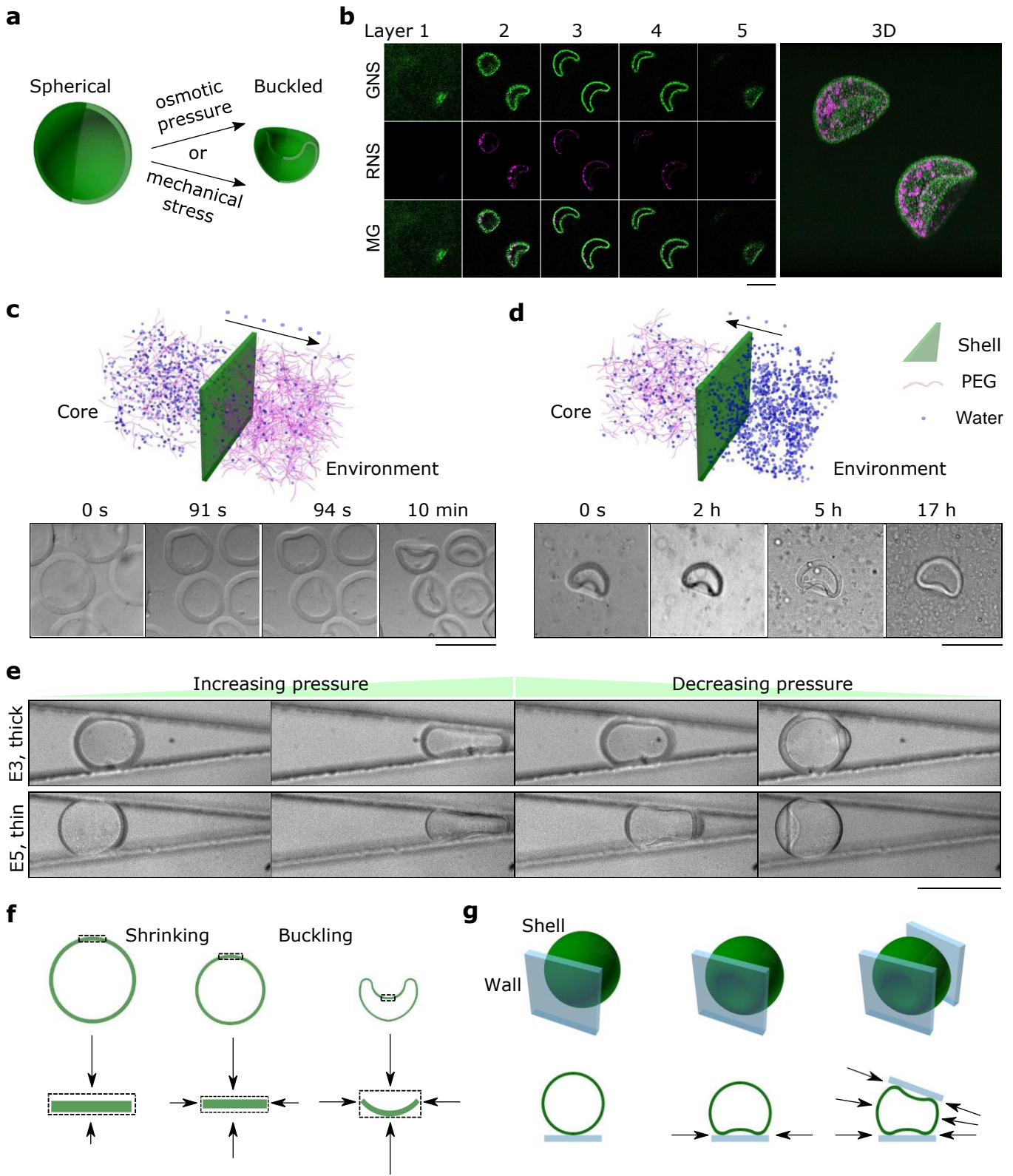

**Fig. 3 The stress-sensitive buckling of microcapsules. a**, Schematic of a cross section of a spherical microcapsule and a buckled microcapsule by osmotic pressure or mechanical pressure. **b**, Confocal imaging of a buckled microcapsule by osmotic pressure (**Movie S7**,**S8**). Scale bar, 100 μm. **c**,**d**, Time-lapse images of the dehydration (buckling, **c**) of microcapsules by osmotic pressure, and the rehydration (recovery, **d**) of a buckled microcapsule of **c** (**Movie S9**,**S10**). Scale bar, 100 μm. Continued on next page.

**e**, The bucking of a thick-shell or thin-shell microcapsule with increasing and decreasing oil pressure in a V-shaped microfluidic channel (**Movie S11**–**S15**). Scale bar, 100 µm. A manuscript was in preparation for this microfluidic device (53). **f,g**, Stress analysis in a buckled microcapsule by osmotic pressure (**f**) in **c**, and by mechanical pressure (**g**) in **e**.

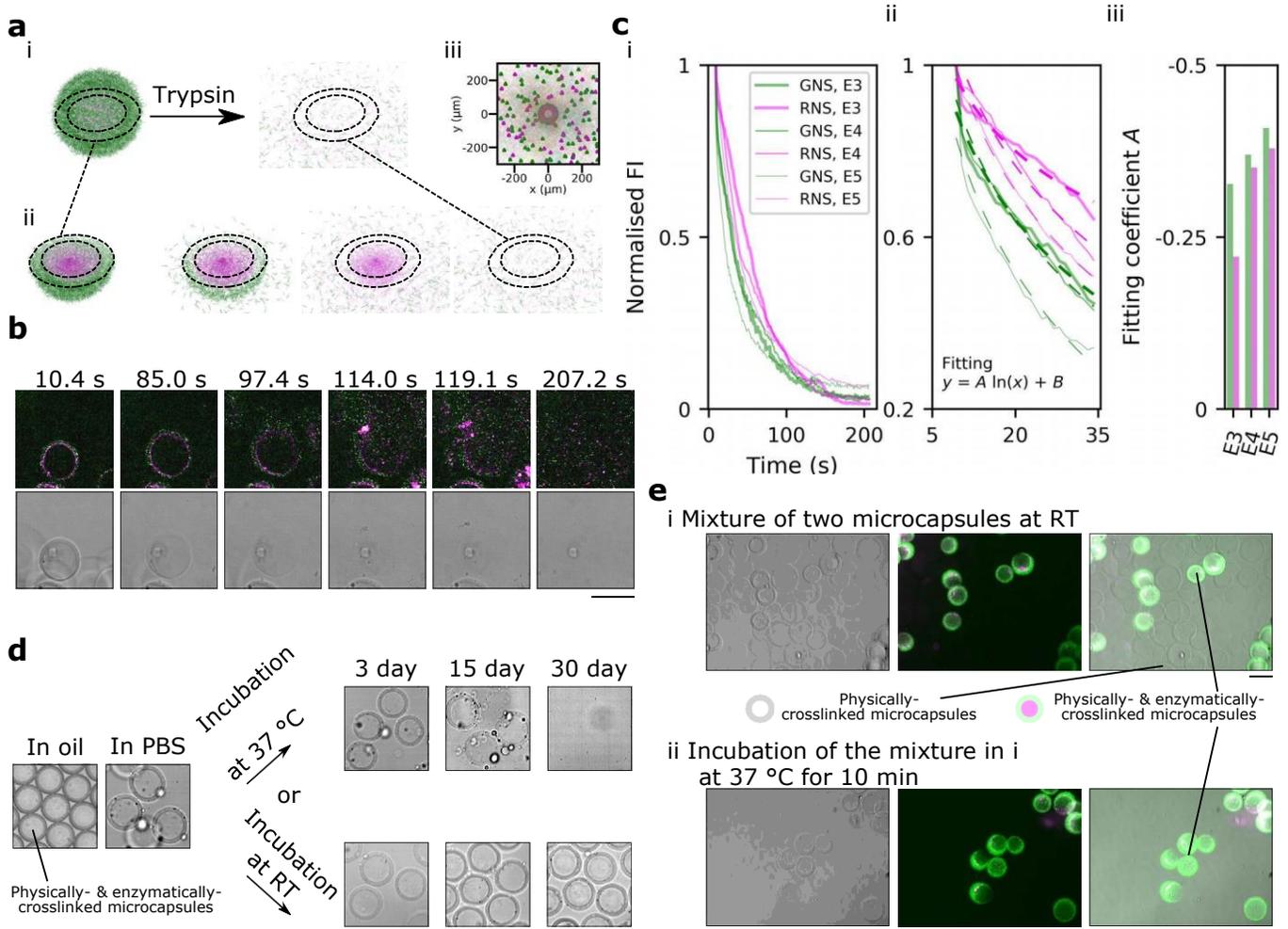

**Fig. 4 Sequential dual release of nanospheres from the microcapsules and the thermostability of the microcapsules a**, Schematics of the sequentially dual release of nanospheres of microcapsules. i, a full microcapsule. ii, The cross section of the microcapsule. The gelatin/PBS and gelatin/PEG boundaries were shown by dashed circles. iii, Simulation (**Supplementary Fig. 15**) of the nanosphere positions and moving paths following Brownian motion, before the shell dissolution of a microcapsule (GNSs, green circles; RNSs, red circles) and after the complete dissolution of the microcapsule (GNSs, green triangles; RNSs, red triangles). **b**, Time-lapsed images (2D confocal imaging) of the release of GNSs from shells and the RNSs from the cores of the microcapsules with medium thickness (E4) during the enzymatic digestion. Scale bar, 100 μm. **c**, i, Sequential dual release of GNSs from shells and RNSs from cores of microcapsules with varying shell thickness. Standard deviation and more details can be found in **Supplementary Fig. 14**, and the sample size of E3–E5 is respectively 126, 107, 108. ii, The beginning of the release curves in i was fitted to $y = A \ln(x) + B$. Coefficient $A$ was shown in bar chart. iii, Coefficient $A$ was shown in bar chart. **d**, Thermostability of physically-crosslinked and enzymatically-crosslinked microcapsules (E3). Scale bar, 100 μm. **e**, i, The mixture of the two kinds of microcapsules was kept at RT. ii, Then the physically-crosslinked microcapsules dissolved after the mixture was incubated at 37 °C for 10 min. Scale bar, 100 μm.

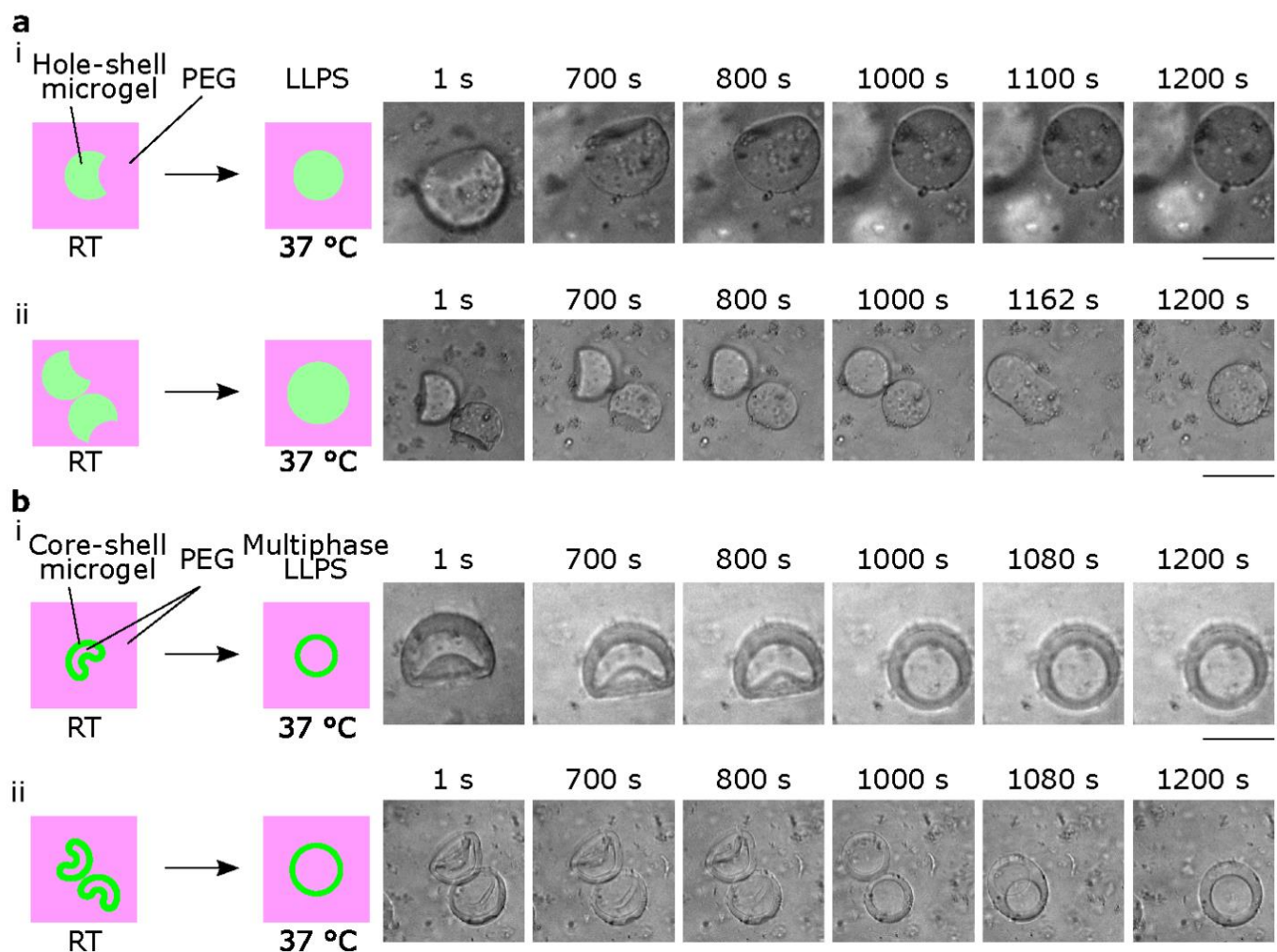

**Fig. 5 All-aqueous LLPS of physically crosslinked microgels in a macromolecular crowder. a**, All-aqueous LLPS. i, Liquefaction of a hole-shell microgel in PEG solution (**Movie S16**). Scale bar, 50 μm. ii, Coalescence of two liquefied hole-shell microgels in PEG solution (**Movie S17**). Scale bar, 100 μm. **b**, All-aqueous multiphase LLPS. i, Liquefaction of a buckled core-shell microgel in PEG solution (**Movie S18**). Scale bar, 50 μm. ii, Coalescence of two liquefied core-shell microgels in PEG solution (**Movie S19**). Scale bar, 100 μm. Finite element simulation can be found in **Supplementary Fig. 20**.

# Supplementary information

# Deformable and robust core-shell protein microcapsules templated by liquid-liquid phase separated microdroplets

Yufan Xu[1], Yi Shen[1], Thomas C. T. Michaels[1,3], Kevin N. Baumann[1,2], Daniele Vigolo[4], Quentin Peter[1], Yuqian Lu[1], Kadi L. Saar[1], Dominic Vella[5], Hongjia Zhu[1], Bing Li[1], He Yang[1,7], Alexander P. M. Guttenplan[1,6], Marc Rodriguez-Garcia[1], David Klenerman[1], and Tuomas P. J. Knowles[1,2]*

[1]Department of Chemistry, University of Cambridge, Cambridge, CB2 1EW, United Kingdom. [2]Cavendish Laboratory, University of Cambridge, Cambridge, CB3 0HE, United Kingdom. [3]Paulson School of Engineering and Applied Sciences, Harvard University, Cambridge, MA 02138, United States. [4]School of Chemical Engineering, University of Birmingham, Edgbaston, Birmingham, B15 2TT, United Kingdom. [5]Mathematical Institute, University of Oxford, Woodstock Rd, Oxford, OX2 6GG, United Kingdom. [6]Department of Pharmacology, University of Cambridge, Cambridge CB2 1PD, United Kingdom. [7]School of Mechanical Engineering, Hangzhou Dianzi University, No. 1158, No. 2 Street, Jianggan District, Hangzhou 310018, P.R. China. *e-mail: tpjk2@cam.ac.uk

# Supplementary information

Common core-shell micro constructs mainly exist in the forms of water/oil/water emulsion or oil/water/oil emulsion (1; 2, SI ref), mono-material core-shell microgels (3, SI ref), and artificial lipid vesicles (4–6, SI ref). Among these constructs, gel microcapsules are mechanically robust and are promising to be used as building blocks for complex structures. Existing microfluidic methods to make microcapsules mainly include two-step formation or the re-injection of microgels into a second microfluidic chip (7; 8, SI ref) and the use of 3D capillary microfluidics (9; 10, SI ref). However, the re-injection of solid microgels into a second chip does not usually have high yield of monodisperse microgels with one core in one shell, which is limited by the sychronisation of the two chips; and the assembly of 3D capillary microfluidics could be time consuming and experience-based. In contrast, a two-dimensional (2D) one-step microfluidic chip based on the aqueous two-phase system would provide a robust and simple method for the formation of core-shell microgels, and could also avoid the problems such as blockage of channels, low yield rates of microgels, and the batch difference of microgels produced in manually-assembled capillaries.

One major reason to use collagen-based materials is that artificial extracellular environment requires biocompatibility, bioactivity, and biomimicry of the biomaterials for regenerative medicine. Collagen is a major component of the human extracellular matrix, and the precise and controllable fabrication of collagen at micron-scale is essential for delicate studies of tissue or organ regeneration, disease models, and protein-based bio reactors. Precursor molecule gelatin could address the engineering problems of native collagen such as its lower solubility at physiological pH (11, SI ref). Gelatin, namely hydrolysed collagen, inherts the tripeptide Arg-Gly-Asp motif of native collagen, and is degradable with matrix metallopeptidases (12; 13, SI ref). Microfluidic techniques offer an approach to manipulate the reconstruction of gelatin with the peculiar physicochemical gelation properties for 3D architectures for cell-related studies in a spatial and temporal manner (14, SI ref).

Enzymatically-crosslinked gelatin untangled the problems of crosslinking strategies with extreme conditions such as aldehydes or UV, and proposed a relatively mild crosslinking approach to gelate hydrogels (11; 15, SI ref). Previous study shows the formation of gelatin microgels with controlled radial density through physical and enzymatic crosslinking of gelatin and microfluidic mixing (11, SI ref). However, the solid microgels do not possess real core-shell structures, because a mono-material system of gelatin was used as the mainly body of the microgels. Solid microgels might encounter some problems such as slow exchange of nutrition and metabolic wastes between the cores and the environments. In contrast, microcapsules could provide hollow or core-shell structures where protein shells have faster interactions with the environment. Compared to the solid microgels which have relatively inhomogeneous or anisotropic structures, the microcapsules could provide even more sharply different mechanical and componental properties of the cores and shells and thus can be further used as heterogeneous micro reactors for bio-related studies. A way to produce microcapsules, is to use dual-material systems which are immiscible or have various gelation characteristics to ensure that the phase equilibrium could be achieved in the microgels.

Microfluidic channels could be blocked by the precipitates or aggregates caused by the direct contact of the PEG and the transglutaminase solutions, so the gelatin solution was injected between PEG and transglutaminase solutions on chip for the formation of physically-crosslinked and enzymatically-crosslinked gelatin microgels (**Main text Fig. 1a** and **Supplementary Fig. 2**) (16, SI ref).

# Supplementary information

**Supplementary Table 1** All-aqueous systems showing liquid-liquid phase separation

| Phase 1 | Phase 2 | References |
|---|---|---|
| Fused in Sarcoma protein | Salt solution (concentration dependant) | (17, SI ref) |
| Ded1 | PIPES buffer (pH dependent) | (17, SI ref) |
| Annexin A11 | Dextran solution | (17, SI ref) |
| zFF | Dextran solution | (17, SI ref) |
| Silk | Dextran solution | (17, SI ref) |
| Dextran solution | PEG solution | (18–22, SI ref) |
| Dextran + PAAm solution | PEG solution | (19, SI ref) |
| Dextran + PAA solution | PEG solution | (19, SI ref) |
| Gelatin solution | PEG solution | (23; 24, SI ref) |
| Gelatin solution | Dextran solution | (25, SI ref) |
| GelMA solution | PEG solution | (26, SI ref) |

PEG: polyethylene glycol; PAAm: polyacrylamide; PAA: poly(acrylic acid); GelMA: gelatin-methacryloyl; zFF: carboxybenzyl (Z)-protected diphenylalanine; PIPES: piperazine-N,N-bis(2-ethanesulfonic acid).

# Supplementary information

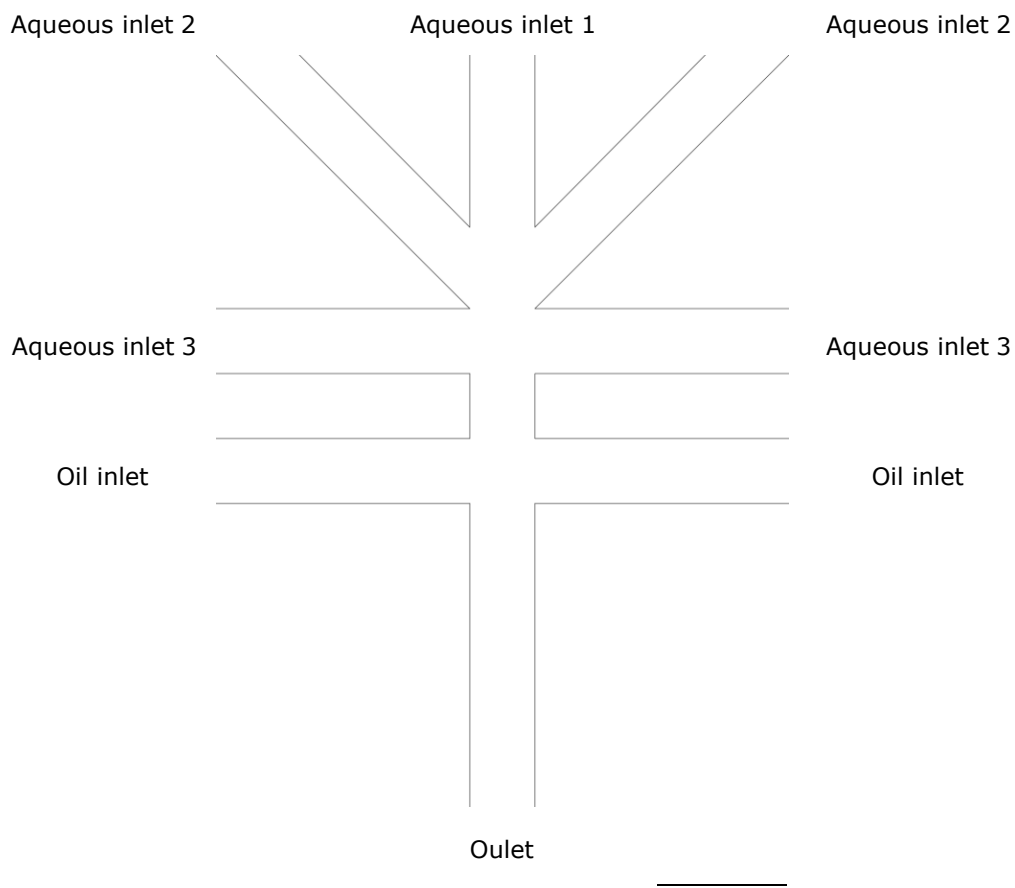

**Supplementary Fig. 1  AutoCAD design of a V-shaped junction.** There are three aqueous inlets, one oil inlet, and one outlet. Scale bar, 200 μm.

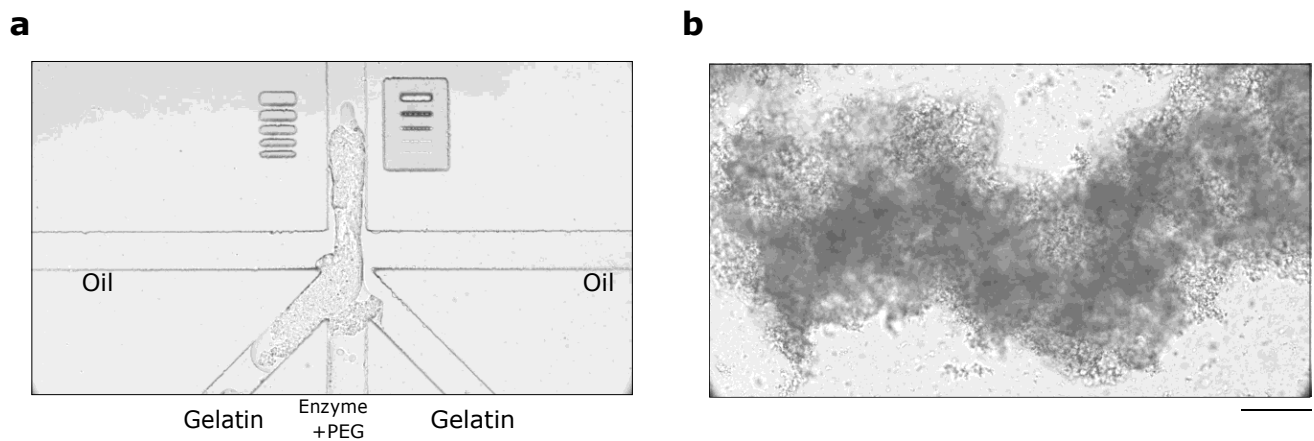

**Supplementary Fig. 2 Blockage of the channels. a**, Blockage in a three-inlet V-shaped flow-focusing droplet-making device. Scale bar, 200 µm. **b**, Precipitates formed in the mixture of transglutaminase solution and PEG solution. Scale bar, 200 µm.

# Supplementary information

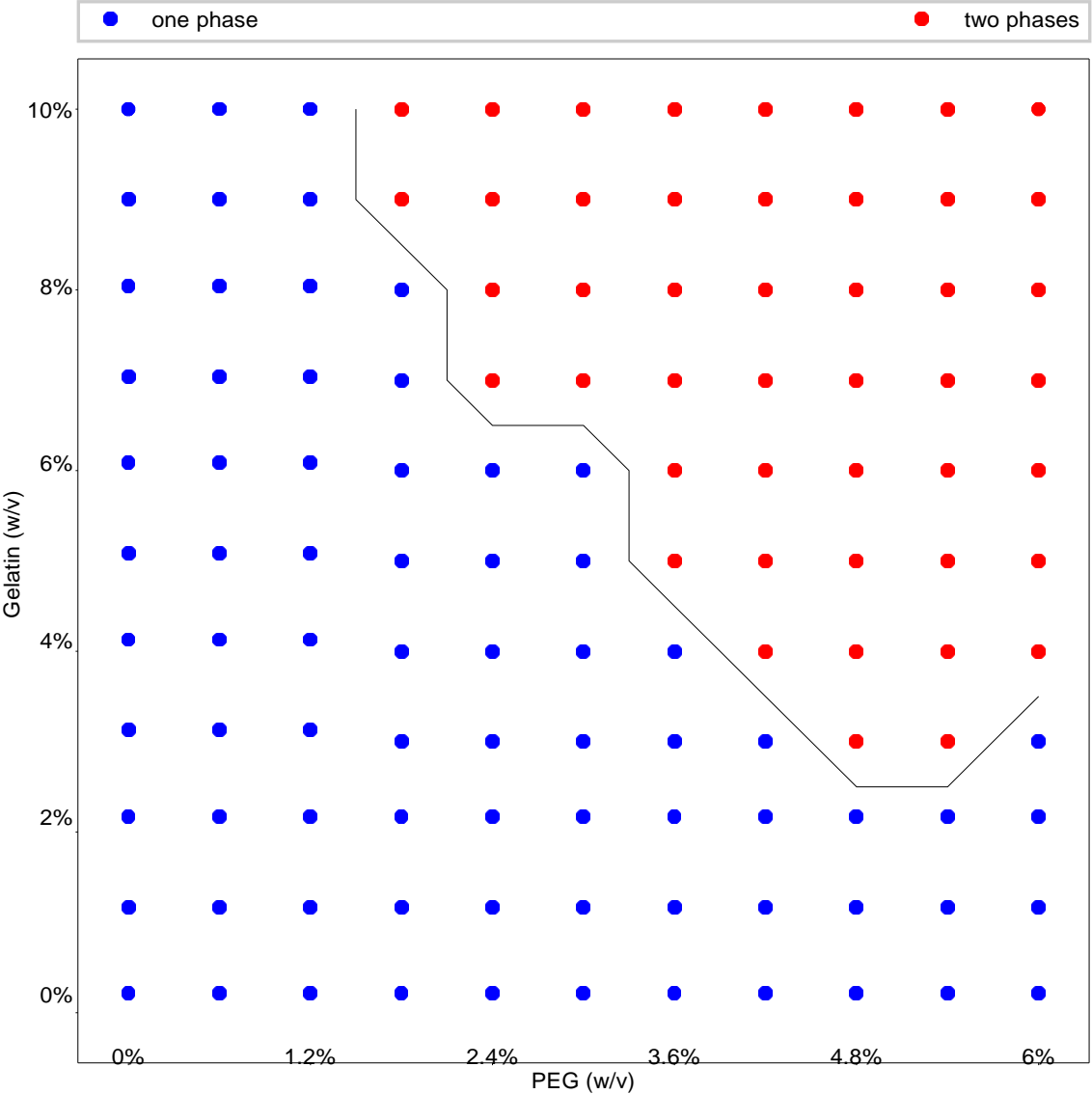

**Supplementary Fig. 3  Phase diagram of the mixing of gelatin solution and PEG solution.** The gelatin solution and PEG solution of different concentrations were mixed at 37 °C.

# Supplementary information

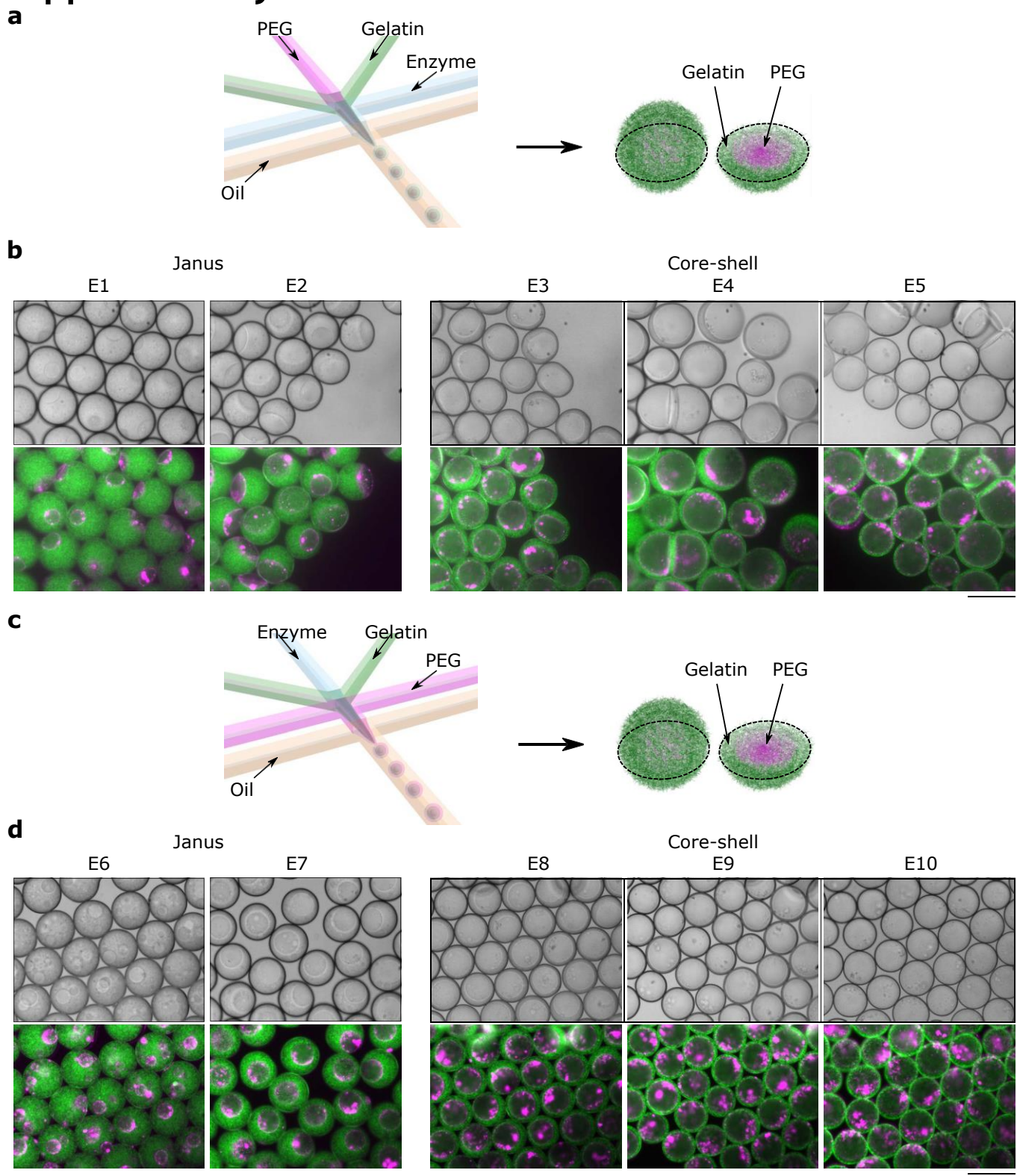

**Supplementary Fig. 4 Two geometries for making microgels. a,b,** A PEG-in-gelatin geometry during droplet generation. The microgels (**b**) are in oil. Scale bar, 100 µm. **c,d,** A gelatin-in-PEG geometry during droplet generation. The microgels (**d**) are in oil. Scale bar, 100 µm.

# Supplementary information

## Theory of the system surface energy

Consider droplets of fixed total volume $V_{tot}$ consisting of two phase-separated components, $A$ and $B$. The total volume of the droplet is fixed, but the volume fractions of $A$ and $B$ can change; let $x_A$ and $x_B = 1 - x_A$ denote the volume fractions of components $A$ and $B$. The volumes of phases $A$ and $B$ are thus

$$V_A = x_A V_{tot} \quad \Rightarrow \quad V_B = (1 - x_A) V_{tot}. \tag{1}$$

Let $\gamma_A$, $\gamma_B$, and $\gamma_{AB}$ be respectively the interfacial tension of component $A$ with the surrounding oil, the interfacial tension of $B$ with the oil, and the interfacial tension between phases $A$ and $B$. We are interested in understanding the conditions under which the distribution of $A$ and $B$ inside the droplet leads to a Janus or a core-shell configuration (**Supplementary Fig. 5a**). To do so, we calculate and then compare the surface energies for these configurations.

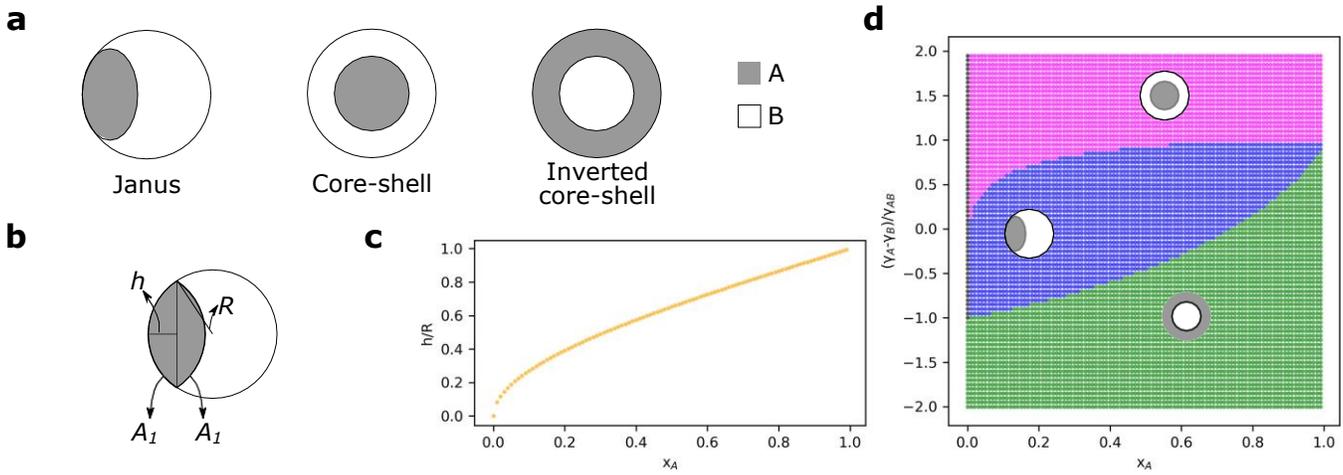

**Supplementary Fig. 5 a**, Possible droplet configurations in the model: Janus, core-shell and inverted core-shell. **b** and **c** Geometry of Janus configuration and dependence of h/R (cap height / sphere radius) on $x_A$. **d**, Phase diagram as a function of $x_A$ and $(\gamma_A - \gamma_B)/\gamma_{AB}$ illustrating the minimum energy configurations.

## Energy of core-shell configuration

In this configuration, a spherical core of component $A$ is surrounded by a spherical shell of component $B$. The radius $R_c$ of the inner core is obtained from the volume of phase $A$ as

$$V_A = x_A V_{tot} \quad \Rightarrow \quad R_c = x_A^{1/3} R, \tag{2}$$

where $R$ is the radius of the entire droplet, $V_{tot} = 4\pi R^3/3$. The contact area between components $A$ and $B$ is thus given by $A_c = 4\pi R^2 x_A^{2/3}$ and the total energy of the core-shell configuration is

$$E_{core-shell} = 4\pi R^2 (\gamma_{AB} x_A^{2/3} + \gamma_B). \tag{3}$$

# Supplementary information

## Energy of inverted core-shell configuration

We can carry out an analogous calculation for the inverted core-shell configuration, where a spherical core of component $B$ is surrounded by a spherical shell of component $A$. In this case, the radius $R_c$ of the inner core is

$$V_B = (1 - x_A)V_{tot} \quad \Rightarrow \quad R_c = (1 - x_A)^{1/3}R. \tag{4}$$

The surface area in contact with $A$ and $B$ is thus $A_c = 4\pi R^2 (1 - x_A)^{2/3}$ and the total energy of this inverted core-shell configuration is

$$E_{inverted} = 4\pi R^2(\gamma_{AB}(1 - x_A)^{2/3} + \gamma_A). \tag{5}$$

## Energy of Janus configuration

The energy of Janus configuration is given by

$$E_{Janus} = \gamma_A A_1 + \gamma_{AB} A_1 + \gamma_B(4\pi R^2 - A_1), \tag{6}$$

that is to say

$$E_{Janus} = (\gamma_A + \gamma_{AB} - \gamma_B)A_1 + \gamma_B 4\pi R^2, \tag{7}$$

where $A_1$ (left) is the contact area between $A$ and oil, while $A_1$ (right) is the contact area between $A$ and $B$. $A$ contains two spherical caps of each height $h$ (**Supplementary Fig. 5b**); the surface area of one cap is given by

$$A_1 = 2\pi Rh, \tag{8}$$

and the volume of the cap is given by

$$V_{cap} = \frac{2\pi h^2(3R-h)}{3}. \tag{9}$$

The volume of phase A is known as

$$V_A = 2\,V_{cap}, \tag{10}$$

that is to say

$$\frac{2\pi h^2(3R-h)}{3} = x_A V_{tot} = \frac{x_A 4\pi R^3}{3}, \tag{11}$$

which is simplified as

$$h^3 - 3Rh^2 + 2x_A R^3 = 0, \tag{12}$$

or

$$(\tfrac{h}{R})^3 - 3(\tfrac{h}{R})^2 + 2x_A = 0. \tag{13}$$

Solving this cubic equation (h/R as a variable) is complicated, so numerical solution is obtained, as shown in **Supplementary Fig. 5c**.

# Supplementary information

## Phase diagram

For comparison of the energies of the different configurations, it is convenient to consider rescaled surface energies, which is denoted using a bar and are obtained by subtracting $E_0 = 4\pi R^2 \gamma_B$ and dividing by $4\pi R^2 \gamma_{AB}$. This yields the three rescaled energy configurations:

$$\bar{E}_{core-shell} = x_A{}^{2/3} \tag{14}$$

$$\bar{E}_{inverted} = (1 - x_A)^{2/3} + \frac{\gamma_A - \gamma_B}{\gamma_{AB}} \tag{15}$$

$$\bar{E}_{Janus} = \frac{1}{2} \frac{h}{R} \left( \frac{\gamma_A - \gamma_B}{\gamma_{AB}} + 1 \right) \tag{16}$$

We see that these energies depend only on two dimensionless parameters, $x_A$ and $(\gamma_A - \gamma_B)/\gamma_{AB}$. This allows us to set up a phase diagram of the different phases as a function of these two parameters in **Supplementary Fig. 5d**. The variation of PEG to gelatin ratio might lead to the water redistribution in the gelatin and PEG phases due to osmosis, and would thus affect $(\gamma_A - \gamma_B)/\gamma_{AB}$, which has been discussed in the main text. Therefore, $(\gamma_A - \gamma_B)/\gamma_{AB}$ varies with the increasing of $x_A$, and the system tended to transit from Janus to core-shell states for minimum energy.

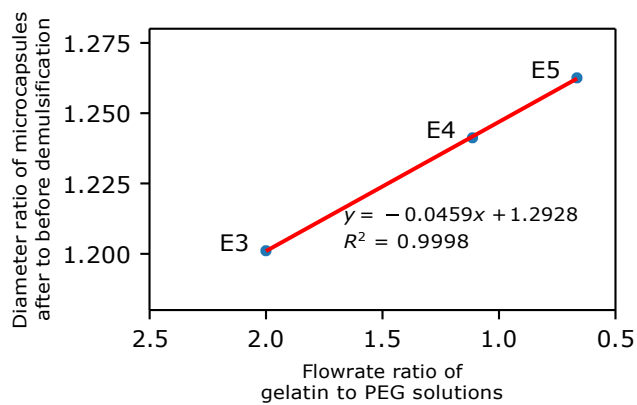

**Supplementary Fig. 6 Swelling of microcapsules during demulsification.** The diameter ratio of the microcapsules after to before demulsification is linear with the flowrate ratio of gelatin to PEG solutions.

# Supplementary information

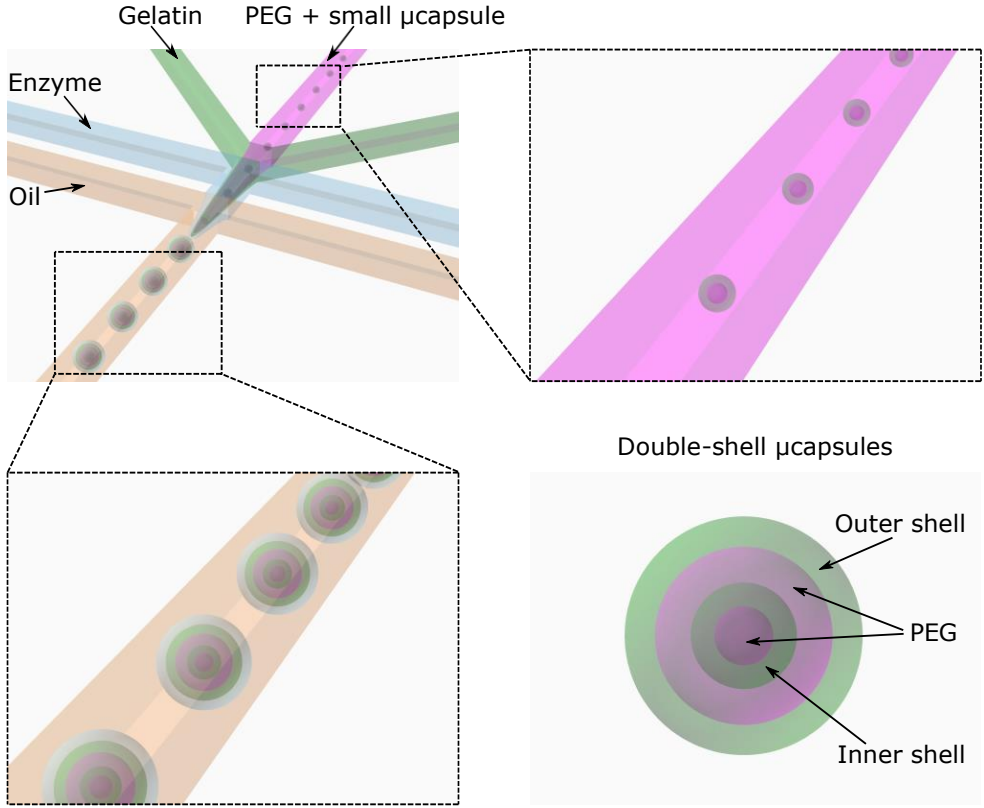

**Supplementary Fig. 7   Formation of double-shell microcapsules.** Small microcapsules are re-injected into big microcapsules to form double-shell microcapsules.

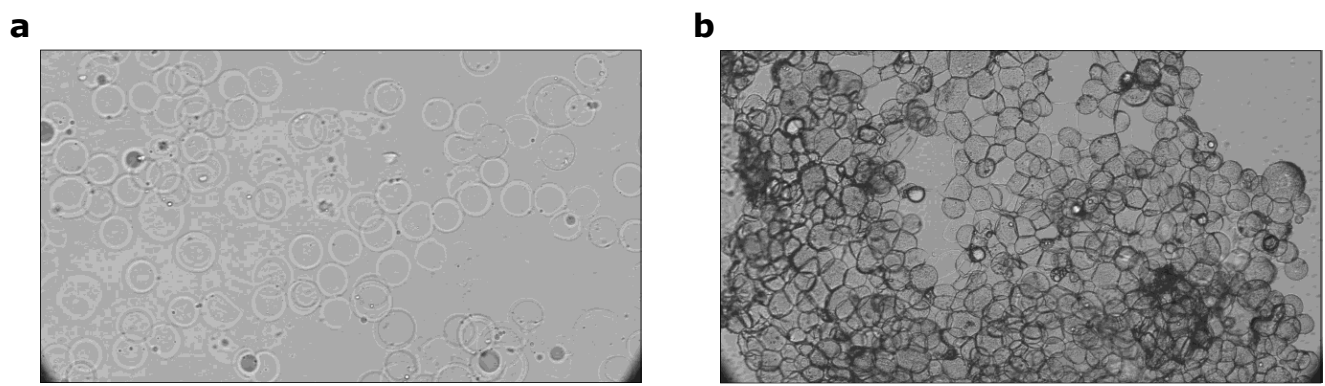

**Supplementary Fig. 8   Dehydration of microgels. a**, Microcapsules were in PBS. Scale bar, 200 μm. **b**, Microcapsules of **a** were dehydrated in ethanol. Scale bar, 200 μm.

# Supplementary information

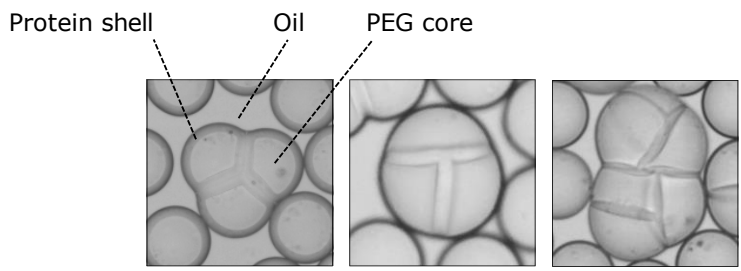

**Supplementary Fig. 9 Multiple microcapsules in a single droplet in oil.** A rotationally symmetric triplet (left), an asymmetric triplet (middle), and a sextuplet (right). Scale bar, 100 μm.

We showed the situation where two microcapsules were encapsulated in a single droplet in oil (**Main text Fig. 2**). Some rotational symmetric or asymmetric structures were found when three or more microcapsules combined (**Supplementary Fig. 9**), which highlighted their possible applications as amorphous and multi-compartmentalised artificial bio reactors (27; 28, SI ref).

# Supplementary information

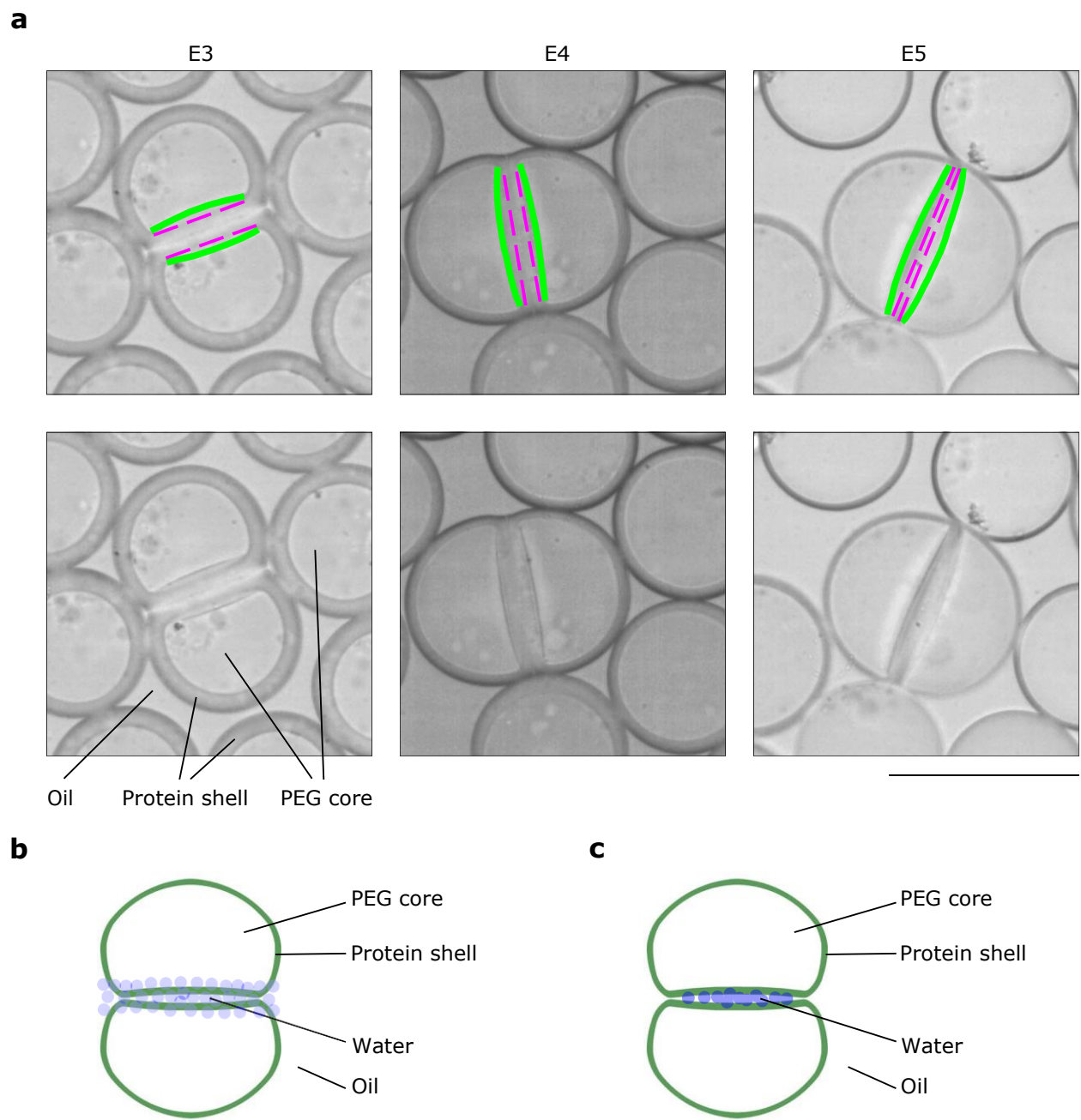

**Supplementary Fig. 10  Buckling of microcapsules when two microcapsules combine in oil. a**, Green curves describe the shell/core interfaces, indicating the buckling of the shells at shell/shell interfaces. Magenta lines indicate the chords of the green curves. Scale bar, 100 μm. **b**,**c**, An explanatory schematic of the form of the water collar squeezed out of the microcapsules by water/oil interfacial tension (**b**) and the form of the water squeezed out of the microcapsules by gel/oil interfacial tension (**c**).

# Supplementary information

## Mechanisms of the phenomenon of two microcapsules combined (encapsulated) in a single oil drop

We here discuss two mechanisms of the deformation of two microcapsules encapsulated in a single droplet in oil (**Main text Fig. 2**). On the one hand, there is a reduction of the volume of water inside the shells, and the water/oil interfacial tension may play a role (**Supplementary Fig. 10b**); this interfacial tension can influence the combining of two microcapsules because the water squeezed out of the microcapsules could form a collar-like bridge around the shell/shell contact area (the capillary pressure in this bridge is negative and hence would act to suck two capsules closer and draw water out, **Supplementary Fig. 10b**). The effect of the capillary pressure was larger on the thinner-shell microcapsules than that on thicker-shell microcapsules, as more water was squeezed out from the cores of thinner-shell microcapsules (**Supplementary Fig. 10b**). On the other hand, there is a reduction of the area of shell/oil interfaces (**Supplementary Fig. 10a**); the minimisation of shell/oil interfacial energy can drive the two microcapsules to touch more closely with some of the water squeezed out existing between the gaps of the two buckled shells (**Supplementary Fig. 10c**).

## Ratio of surface reduction in this phenomenon

The object of two microcapsules combining in oil could be considered as two hemispheres at two sides and a cylinder in the middle, as shown in **Main text Fig. 2a**. Let $W$ and $L$ denote the width and length of this object (**Main text Fig. 2a** and **Supplementary Fig. 10**). The width to length ratio is defined as $m$,

$$m = \frac{W}{L}. \tag{17}$$

The volume of this object is,

$$V_{obj} = 2 \times V_{hemisphere} + V_{cylinder}. \tag{18}$$

$$V_{obj} = 2 \times \frac{1}{2} \times \frac{4\pi}{3} \times (\frac{W}{2})^3 + \pi(\frac{W}{2})^2(L - W) = \frac{m^2(3-m)\pi}{12}L^3. \tag{19}$$

The surface of this object consists of the surfaces of two semi-spherical surfaces and the side surface of the cylinder, which is,

$$S_{obj} = 2 \times S_{hemisphere} + S_{cylinder} = 4\pi(\frac{W}{2})^2 + \pi W(L - W) = m\pi L^2. \tag{20}$$

Obviously,

$$S_{obj} = \left[\frac{144\pi}{m(3-m)^2}V_{obj}{}^2\right]^{1/3}. \tag{21}$$

Based on **Fig. 2c**, $m$ of a combining microcapsule is 0.6788, 0.7572, and 0.7843 respectively for E3, E4, and E5. $V_{obj}$ of a combining microcapsule stays the same for E3, E4, and E5 microcapsules. For two microcapsules which have not combined yet, the gel/oil surface area is,

$$S_{obj_{Eoriginal}} = \left[\frac{144\pi}{1\times(3-1)^2}(\frac{V_{obj}}{2})^2\right]^{1/3}. \tag{22}$$

# Supplementary information

For two E3 microcapsules combining, the gel/oil surface area is,

$$S_{obj_{E3}} = \left[\frac{144\pi}{0.6788\times(3-0.6788)^2}V_{obj}{}^2\right]^{1/3}. \tag{23}$$

For two E4 microcapsules combining, the gel/oil surface area is,

$$S_{obj_{E4}} = \left[\frac{144\pi}{0.7572\times(3-0.7572)^2}V_{obj}{}^2\right]^{1/3}. \tag{24}$$

For two E5 microcapsules combining, the gel/oil surface area is,

$$S_{obj_{E5}} = \left[\frac{144\pi}{0.7843\times(3-0.7843)^2}V_{obj}{}^2\right]^{1/3}. \tag{25}$$

For two microcapsules combining into a bigger sphere hypothetically, the gel/oil surface area is,

$$S_{obj_{Eend}} = \left[\frac{144\pi}{1\times(3-1)^2}V_{obj}{}^2\right]^{1/3}. \tag{26}$$

The surface reduction ratio of $Ei$ is defined as,

$$SRR_{obj_{Ei}} = 1 - \frac{S_{obj_{Ei}}}{S_{obj_{Eorginal}}}. \tag{27}$$

The surface reduction ratio is 0 for two microcapsules which have not combined yet; the surface reduction ratio is 0.1822, 0.1932, and 0.1961 respectively for E3, E4, and E5; the surface reduction ratio is 0.2063 for two microcapsules combining into a bigger sphere hypothetically (**Main text Fig. 2d**).

# Supplementary information

**a**    **b**

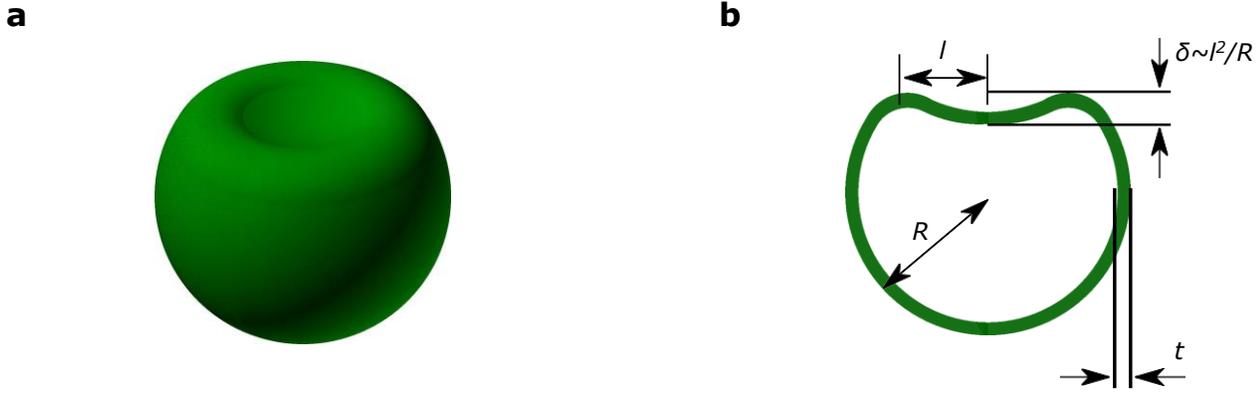

**Supplementary Fig. 11 Schematics of a buckled microcapsule. a**,**b**, 3D (**a**) and cross-section (**b**) schematic of a buckled capsule. $R$, radius of the capsule. $t$, thickness of the capsule. $E$, Young's modulus of the capsule. $l$, radius of the dimple. $\delta$, depth of the dimple.

Here we present a scaling argument for the buckling threshold of the shell under pressure. During the external pressurisation of a shell (**Supplementary Fig. 11**) (29; 30, SI ref), the external pressure induces a compressive stress

$$\sigma \propto pR. \tag{28}$$

Buckling creates a dimple of depth $\delta$ and radius $l \sim (\delta R)^{1/2}$ (**Supplementary Fig. 11b**). The bending energy of the dimple is

$$E_B \sim Et^3 (\tfrac{\delta}{l^2})^2 l^2 \sim Et^3 \tfrac{l^2}{R^2}. \tag{29}$$

The stretching energy (31; 32, SI ref) of the dimple is

$$E_S \sim Et(\tfrac{\delta}{l})^4 l^2 \sim Et \tfrac{l^6}{R^4}. \tag{30}$$

The combination of bending and stretching energy is minimised when

$$l \sim (tR)^{1/2}, \tag{31}$$

which allows us to estimate the compressive stress at the start of buckling as

$$\sigma_c \sim Et\tfrac{\delta^2}{l^2} \sim Et\tfrac{\delta^2}{R^2} \sim E\tfrac{t^2}{R}. \tag{32}$$

Equating $\sigma_c \sim p_c R$, we find the critical external pressure at the start of buckling (29; 30, SI ref)

$$p_c \sim E\tfrac{t^2}{R^2}. \tag{33}$$

# Supplementary information

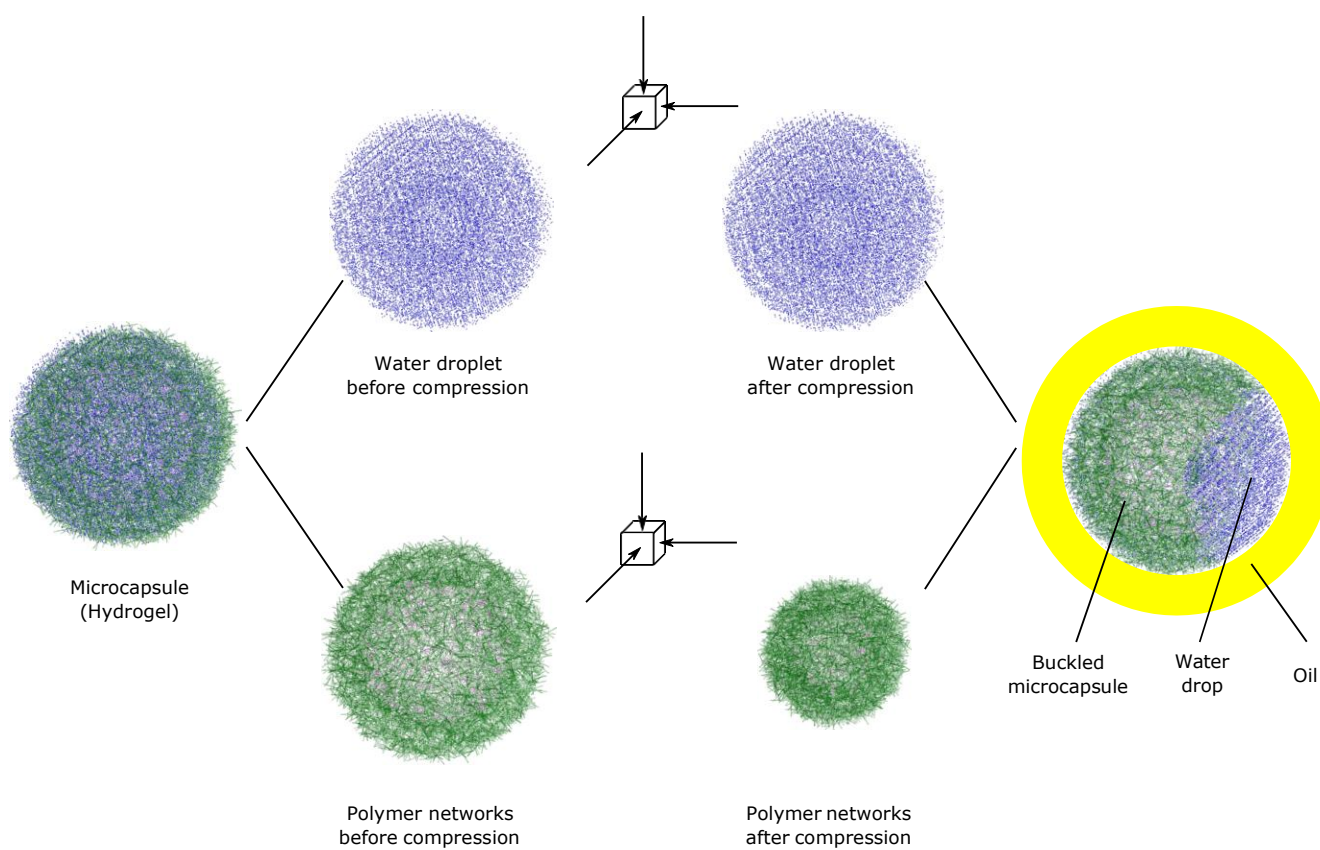

Water droplet
before compression

Water droplet
after compression

Microcapsule
(Hydrogel)

Buckled
microcapsule

Water
drop

Oil

Polymer networks
before compression

Polymer networks
after compression

**Supplementary Fig. 12  Squashing a microcapsule (core-shell microgel) into a channel.** The microgel was considered as two individual parts, an incompressible water droplet, and a compressible sphere of gelatin/PEG polymer networks. Triaxial compressive stresses were shown.

After several loading and unloading cycles,  the water in the PEG-rich core was  squashed out, and thus accumulated outside the microcapsules in oil (**Main text Fig. 3e**). A microcapsule could be considered separately as an incompressible water droplet and a compressible spheroid of gelatin/PEG polymer networks (**Supplementary Fig. 12**). The triaxial compressive stresses condensed the polymer networks. In a capsule in oil, the polymer networks shrank during the squashing,  and then the volume decrement of the polymer was  compensated by  the water that had been squashed out, and thus the microcapsules became flattened or contracted inward (**Main text Fig. 3e**). More water was squashed out of thinner-shell microcapsules (E5) than thicker-shell microcapsules (E3), as obvious buckling was found in thinner-shell microcapsules (E5) (**Main text Fig. 3e** and **Movie S11–S15**). This again indicated that the thicker-shell microcapsules were more mechanically robust (**Supplementary Fig. 11**).

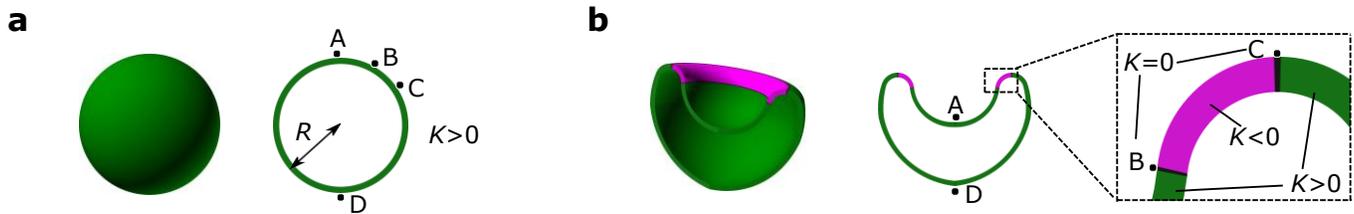

**Supplementary Fig. 13 Distribution of Gaussian curvature $K$ before and after buckling of a spherical capsule. a**, Before buckling, a spherical microcapsule has Gaussian curvature $K = 1/R^2$ constant throughout. **b**, After buckling, the Gaussian curvature is largely preserved (green areas correspond to $K > 0$) except in a narrow 'Pogorelov' ridge (magenta corresponds to $K < 0$, black to $K = 0$).

Changes of Gaussian curvature $K$ of the capsule can only occur by stretching (33, SI ref). Before buckling, for all points on a sphere, $K = 1/R^2 > 0$; after buckling, $K$ remains positive in some areas (green; A, A–B, C–D, and D), but elsewhere $K = 0$ (black; B and C) or $K < 0$ (magenta; B–C) (**Supplementary Fig. 13**). Therefore, even though point A has the largest deformation in terms of the displacement (from hill to valley) during buckling, it is the region B–C that has the highest strain (stretching and compression) after buckling and has highest stress (**Supplementary Fig. 13**) (31; 33, SI ref).

# Supplementary information

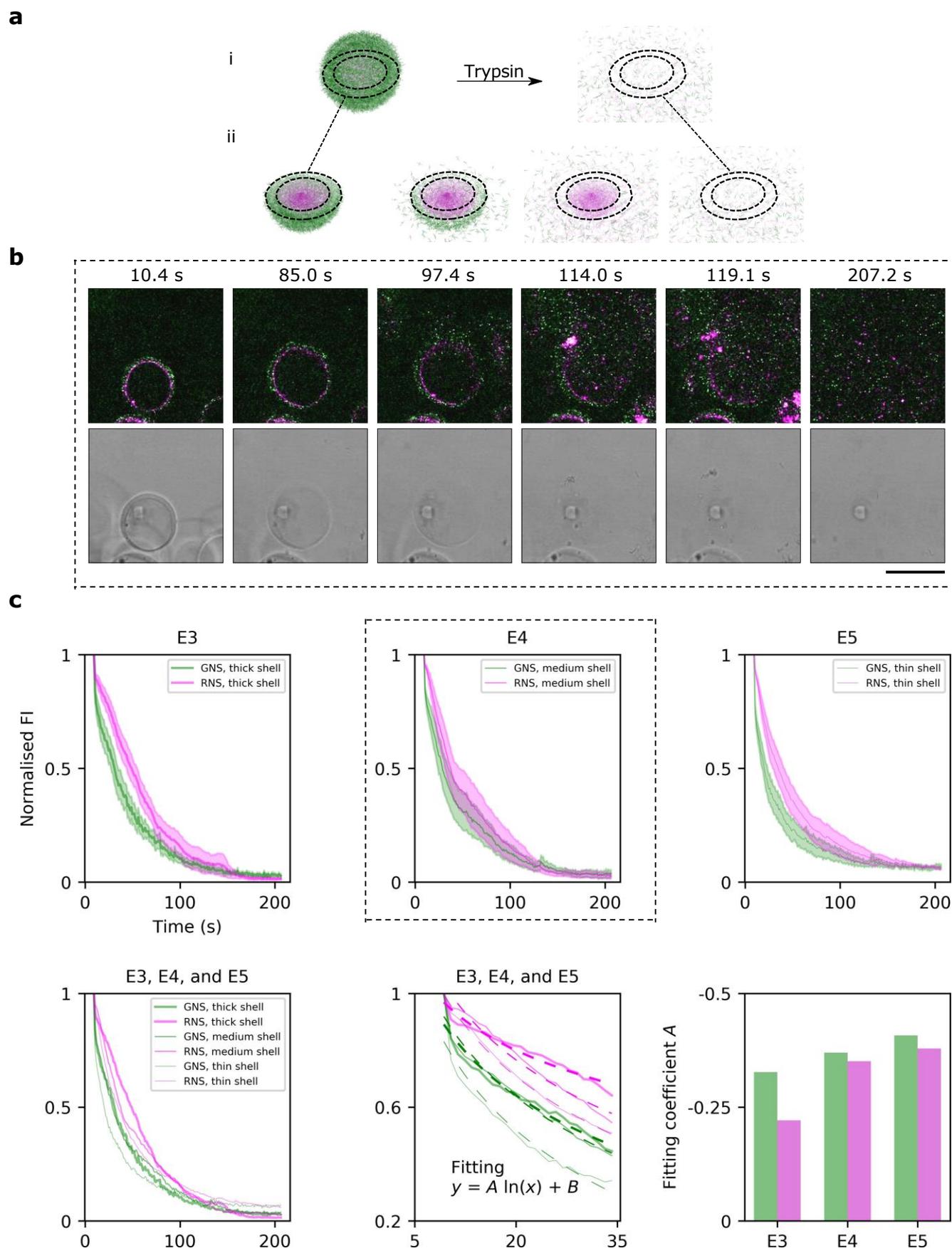

**Supplementary Fig. 14 Sequential dual release of nanospheres from the microcapsules and the thermostability of the microcapsules. a**, Schematics of the successive dual release of nanospheres of microcapsules. i, a full core-shell microgel. Continued on next page.

# Supplementary information

**Supplementary Fig. 14 continues.** ii, the cross section of a core-shell microgel. The boundaries of gelatin/PBS and gelatin/PEG were shown by dashed circles. **b**, Time-lapsed images (2D confocal imaging) of the release of GNSs from shells and the RNSs from the cores of the core-shell protein microgels with medium thickness (E4) during enzymolysis of the shells of microgels. Scale bar, 100 μm. **c**, Sequential dual release of nanospheres of core-shell protein microgels with varying thickness. Standard deviation was shown, and the sample size of E3–E5 is respectively 126, 107, 108. The release curves were fitted to $y = A \ln(x) + B$. Coefficient $A$ was shown in bar chart.

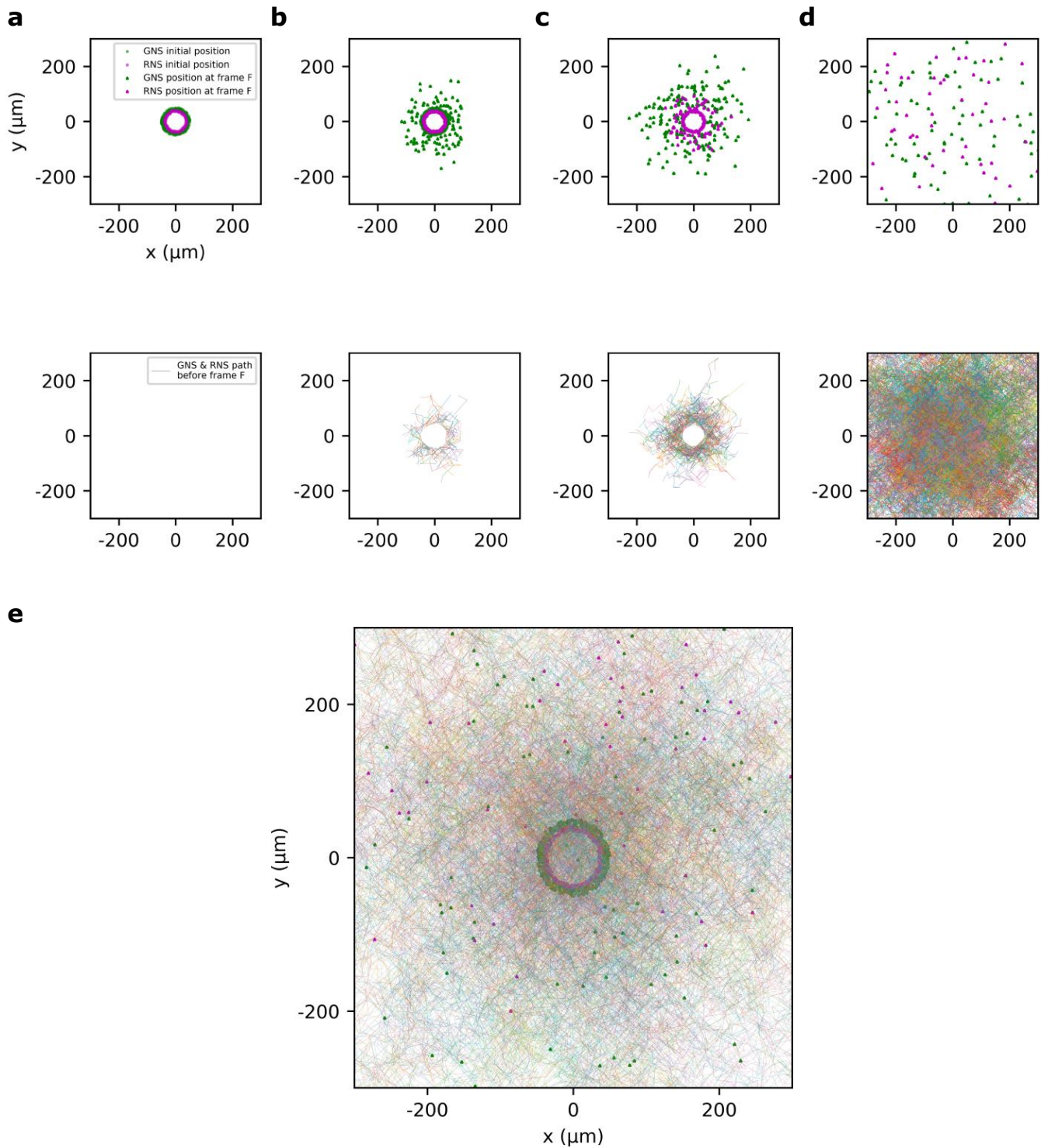

**Supplementary Fig. 15  2D simulation of the sequential release of nanospheres from a microcapsule with the dissolution of the shell. a**,**b**,**c**,**d**, the release of GNSs and RNSs from the a microcapsule at frame F. F = 0 (**a**),  F = 5 (**b**),  F = 13 (**c**),  and F = 200 (**d**).  Top images, the positions of nanospheres at frame F. Bottom images, the path of nanospheres following Brownian motion before Frame F. **a**, Before the dissolution of the shell. **b**, At the start of the GNS release. **c**, At the start of the RNS release. **d**, The uniform distribution of the GNSs and RNSs some time after the complete dissolution and collapse of the microcapsule. **e**, An overlap of F = 0 (**a**) and F = 200 (**d**).

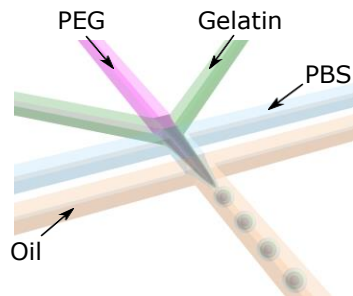

**Supplementary Fig. 16   Making physically-crosslinked microcapsules.** PEG solution, gelatin solution, PBS, and oil were used.

# Supplementary information

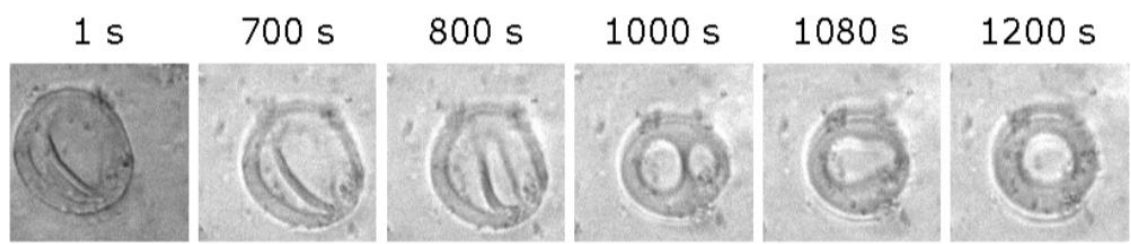

**Supplementary** Fig. 17 Another example of the liquefaction of a physically crosslinked buckled core-shell microgel in PEG solution during heating. Microgels were mixed with PEG solution at RT, and then heated to 37 ˚C. Scale bar, 50 μm.

# Supplementary information

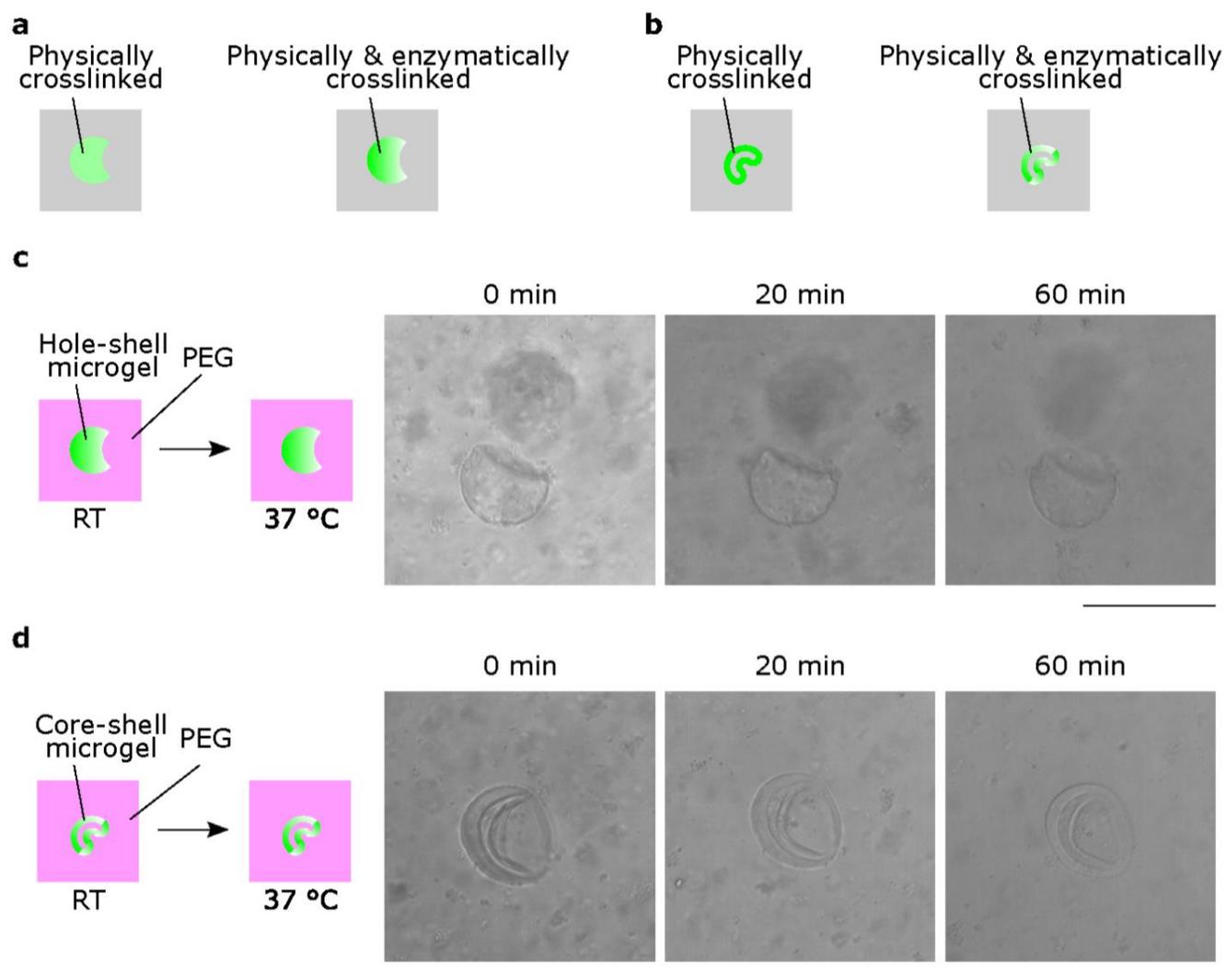

**Supplementary** Fig. 18 No LLPS was observed in the mixture of enzymatically crosslinked microgels and PEG solution during heating. **a**,**b**, Schematics of the enzymatic crosslinking (by transglutaminase) of physically crosslinked hole-shell (**a**) and buckled core-shell (**b**) microgels. **c**,**d**, The enzymatically crosslinked hole-shell (**c**) and buckled core-shell (**d**) microgels remained intact morphologically in PEG solution during heating. Scale bar, 100 μm.

# Supplementary information

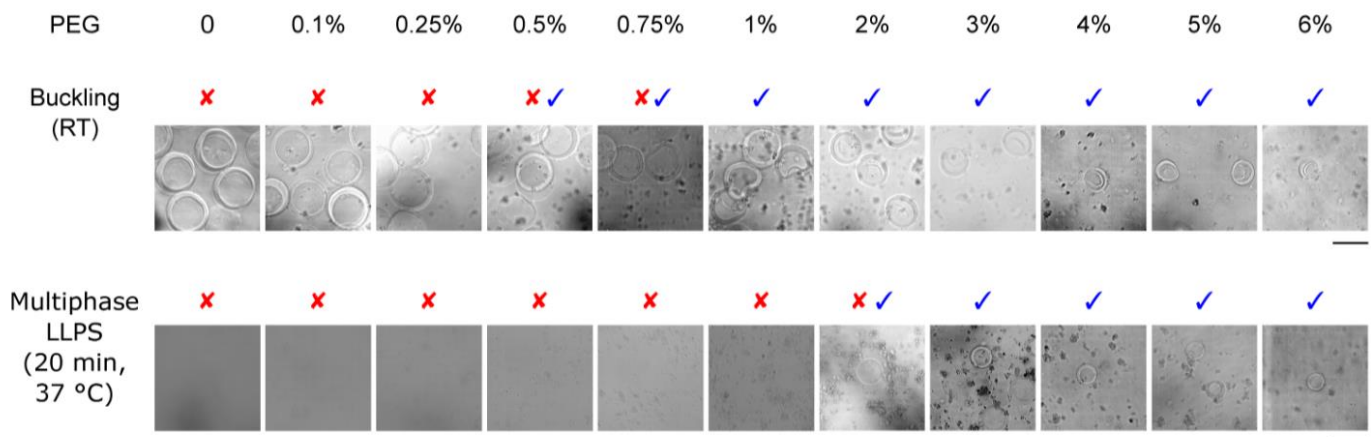

**Supplementary Fig. 19 Buckling of and multiphase LLPS from core-shell microgels in PEG solution of varying PEG concentration.** The buckling of physically crosslinked microgels took place at RT. The multiphase LLPS took place after incubation of physically crosslinked microgels at 37 ˚C for 20 min. Scale bar, 100 µm.

# Supplementary information

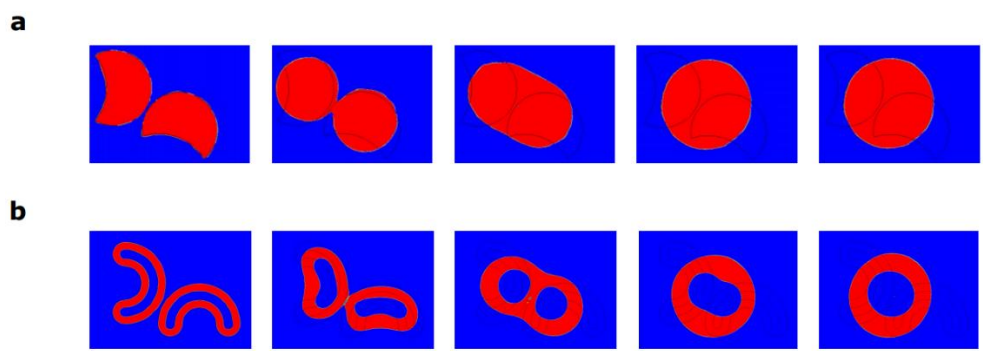

**Supplementary Fig. 20 Finite element simulation of the coalescence of droplets in all-aqueous LLPS and multiphase-LLPS systems. a**, Coalescence of hole-shell structures in all-aqueous LLPS systems. **b**, Coalescence of core-shell structures in all-aqueous multiphase-LLPS systems.

# Supplementary information

## List of movie legends

**Movie S01:** Formation of microdroplets at the flow-focusing junction of the microfluidic chip.

**Movie S02:** Microdroplets near the outlet of the microfluidic chip.

**Movie S03:** Z-stack imaging of the hole-shell microgels.

**Movie S04:** 3D reconstruction of the hole-shell microgels in **Movie S03**.

**Movie S05:** Z-stack imaging of the microcapsules (core-shell microgels).

**Movie S06:** 3D reconstruction of the microcapsules in **Movie S05**.

**Movie S07:** Z-stack imaging of the microcapsules buckled by osmosis.

**Movie S08:** 3D reconstruction of the microcapsules buckled by osmosis in **Movie S07**.

**Movie S09:** Video of images of the buckling (dehydration) of microcapsules by osmosis. Duration = 10 min.

**Movie S10:** Video of images of the recovery (rehydration) of microcapsules by osmosis. Duration = 17 h.

**Movie S11:** Deformation of a thicker-shell microcapsule (E3) indicated by the formation of a tiny water drop between the gel and the oil.

**Movie S12:** Further deformation of a thicker-shell microcapsule (E3) indicated by the formation of a tiny water drop between the gel and the oil after **Movie S11** .

**Movie S13:** Deformation of a thinner-shell microcapsule (E5) indicated by the formation of a water layer between the gel and the oil.

**Movie S14:** Buckling of a thinner-shell microcapsule (E5) after **Movie S13**.

**Movie S15:** The remaining of the buckling of a thinner-shell microcapsule (E5) after the decrease of oil pressure after **Movie  S14**.

Movie S16: Liquefaction of physically crosslinked hole-shell microgels in PEG solution during heating up. Duration = 1326 s.

Movie S17: Coalescing of liquefied hole-shell microgels in PEG solution during heating up. Duration = 1520 s.

Movie S18: Liquefaction of physically crosslinked buckled core-shell microgels in PEG solution during heating up. Duration = 1301 s.

Movie S19: Coalescing of liquefied core-shell microgels in PEG solution during heating up. Duration = 1485 s.

# Supplementary references